\documentclass[
  a4paper,
  10pt,
  numbers=noenddot,    
  parskip=half         
]{scrartcl}

\usepackage[T1]{fontenc}
\usepackage[utf8]{inputenc}
\usepackage{microtype}

\usepackage{mathptmx}
\usepackage[scaled=.90]{helvet}  

\usepackage[
  a4paper,
  top=25mm, bottom=25mm,
  left=20mm, right=20mm
]{geometry}

\setkomafont{sectioning}{\normalfont\sffamily\bfseries}
\setkomafont{title}{\normalfont\sffamily\bfseries}
\setkomafont{author}{\normalfont\small}
\setkomafont{date}{\normalfont\small}

\usepackage{natbib}
\usepackage{hhline}

\usepackage{multirow}
\usepackage{mathtools}

\usepackage[table]{xcolor}
\usepackage{graphicx}
\usepackage{tcolorbox}
\usepackage{xifthen} 

\usepackage{relsize}
\usepackage{hyperref}
\usepackage{balance}
\usepackage{import}
\usepackage{cleveref}
\usepackage{booktabs}

\usepackage{caption}
\usepackage{fontawesome5}
\graphicspath{{./images/}}
\usepackage{pdflscape}

\usepackage{enumitem} 

\usepackage{tikz}
\usepackage{varwidth}

\usepackage[framemethod=TikZ]{mdframed}

\usepackage[shortcuts]{extdash} 

\begin{document}

\usetikzlibrary{patterns}


\makeatletter
\newcommand{\review}[4]{}
\makeatother

\newenvironment{textboxtemplate}[3][]{%
  \par\nobreak\penalty0
  \vtop\bgroup%
  \ifstrempty{#1}
    {\mdfsetup{frametitle={}}}%
        {%
            \mdfsetup{innerleftmargin=10pt, innerrightmargin=10pt,
                frametitle={%
                    \tikz[baseline=(current bounding box.east),
                    outer sep=0pt]
                    \node[anchor=east,rectangle,draw=black!75,fill=#3]
                    {\textcolor{black}{\begin{varwidth}{\linewidth-10pt}~\centering#1\end{varwidth}~}};
                }%
            }%
    }%
  \mdfsetup{innertopmargin=4pt,innerbottommargin=5pt, linecolor=black!75,%
      linewidth=.5pt,topline=true,roundcorner=5pt,skipabove=5pt,
    frametitleaboveskip=\dimexpr-\ht\strutbox\relax}
  \begin{mdframed}[]\relax%
  \label{#2}%
  \setlength{\parindent}{0pt}
}
{%
  \end{mdframed}%
  \par\xdef\tpd{\the\prevdepth}\egroup
  \prevdepth=\tpd%
  }

\newenvironment{textbox}[2][]{%
    \begin{textboxtemplate}[#1]{#2}{tablebg}
  }{
    \end{textboxtemplate}
}

\newenvironment{interviewbox}[2]{
    \begin{textboxtemplate}[#1]{interviewexcerpt:#2}{tablebg}
    }{
\end{textboxtemplate}
}

\newcounter{hyp}
\newenvironment{hypothesisbox}{
    \refstepcounter{hyp}%
    \begin{textboxtemplate}[\faIcon{lightbulb} Hypothesis {\thehyp}]{hyp:\thehyp}{tablebg}
    }{
\end{textboxtemplate}}

\newenvironment{questionbox}[2]{
  \begin{textboxtemplate}[\faIcon{question-circle} RQ{#2}: {#1}]{answer:#2}{tablebg}
    }{
    \end{textboxtemplate}}

\definecolor{pastelgreen}{HTML}{ccebc5}
\definecolor{pastelyellow}{HTML}{fed9a6}
\definecolor{pastelred}{HTML}{fbb4ae}
\definecolor{pastelrose}{HTML}{fddaec}
\definecolor{pastelgray}{HTML}{f2f2f2}
\definecolor{tablebg}{HTML}{f0f0f0}

\definecolor{firefoxrose}{HTML}{de0f63}

\newcommand{\stackoverflow}{Stack Overflow}

 \newcommand{\markasks}[1]{{\color{firefoxrose} Question from Mark: #1}}
 \newcommand{\andy}[1]{{\color{orange} Andy: #1}}
 \newcommand{\rashina}[1]{{\color{red} Rashina: #1}}
 \newcommand{\td}[1]{{\color{firefoxrose} #1}} 

\newcommand{\markmark}[1]{}

\newcommand{\reviewmark}[1]{#1} 
\newcommand{\reviewmarktwo}[1]{#1} 

\newcommand{\loabw}[2][to do: wording]{{\color{red} #2 (#1)}}

\newcommand*\circled[2][tablebg]{\tikz[baseline=(char.base)]{
\node[minimum size=\baselineskip,shape=circle,draw,inner sep=1pt,font=\footnotesize,fill=#1,] (char) {#2};}}

\newcommand\revised[2]{#1} 
\newcommand\revisedtwo[1]{#1} 

\newcommand\freefootnote[1]{%
  \begingroup
  \renewcommand\thefootnote{}\footnote{#1}%
  \addtocounter{footnote}{-1}%
  \endgroup
}

\newcounter{subj}
\refstepcounter{subj}
\makeatletter
\newcommand\subj[3]{%
    \phantomsection%
{\texttt{#3}}\def\@currentlabel{\unexpanded{#3}}\label{subj:#2}}%
\makeatother

\newcommand\interviewq[3]{%
    \begin{interviewbox}{\faIcon{comments} Interview~\ref{subj:#2} at {#3}}{}
        #1
    \end{interviewbox}}

\newcommand\code[1]{\textit{\##1}}

\newcommand\interviewer[1]{{\footnotesize\faIcon{microphone}}:~``\textit{#1}''}
\newcommand\intervieweenoref[1]{{\footnotesize\faIcon{comment}}~``\textit{#1}''}
\newcommand\interviewee[3]{\intervieweenoref{#1} (\ref{subj:#2} at #3)}

\makeatletter
\newcounter{evidencecounter}
\newcommand\ievidence[4]{%
\stepcounter{evidencecounter}%
\phantomsection%
\noindent%
{\footnotesize\textbf{\theevidencecounter.:}} #1%
\def\@currentlabel{{\theevidencecounter}}%
\label{evidence:#2:#3}}%
\makeatother

\newcommand{\groundedin}[2]{{\footnotesize{\textsuperscript{(\ref{evidence:#1:#2})}}}}
\newcommand{\groundedintwo}[4]{{\footnotesize{\textsuperscript{(\ref{evidence:#1:#2},\ref{evidence:#3:#4})}}}}
\newcommand{\groundedinthree}[6]{{\footnotesize{\textsuperscript{(\ref{evidence:#1:#2},\ref{evidence:#3:#4},\ref{evidence:#5:#6})}}}}
\newcommand{\groundedinfour}[8]{{\footnotesize{\textsuperscript{(\ref{evidence:#1:#2},\ref{evidence:#3:#4},\ref{evidence:#5:#6},\ref{evidence:#7:#8})}}}}

\newcounter{insight}
\newcommand\insight[1]{%
    \refstepcounter{insight}%
    \begin{textbox}[]{insight:\theinsight}
        #1
    \end{textbox}
}

\newcommand\subcategory[1]{\textbf{#1}}

\newcommand\weakevidence[0]{{\small\faIcon{user}}}
\newcommand\evidence[0]{\faIcon{user-friends}}
\newcommand\strongevidence[0]{\faIcon{users}}

\newcommand\weblink[2]{\href{#1}{\faIcon{link}~#2}}

\newcommand{\RN}[1]{
  \textup{\uppercase\expandafter{\romannumeral#1}}%
}

\newcommand{\paragraphlf}[1]{\paragraph{#1}\mbox{}\\}

\newcommand{\interviewdetails}[1]{#1}

\title{%
On the Emergence of Testing Strategies:}

\subtitle{%
A Socio-technical Grounded Theory}


\author{%
  Mark Swillus\textsuperscript{1} \and %
  Rashina Hoda\textsuperscript{2} \and %
Andy Zaidman\textsuperscript{1}}

\date{
\textsuperscript{1}Delft University of Technology, Delft, The Netherlands\\
\textsuperscript{2}Monash University, Melbourne, Australia\\
\small Correspondence: \texttt{m.swillus@tudelft.nl}
}


\maketitle
\begin{abstract}
Software testing is crucial
for ensuring software quality,
yet developers' engagement with it varies widely.
Identifying the
technical, organizational and social factors
that lead to differences in engagement
is required to remove barriers and utilize enablers for testing.
\revisedtwo{While much research emphasizes the usefulness
of software testing approaches and technical solutions,
less is known about why developers do (not) test.}
\revisedtwo{
This study investigates the
first-hand experience of developers
with software testing.
The study illuminates
how developers' opinions about testing
and their testing behavior changes.
Through analysis of personal evolutions of practice,
we explore \textit{when} and \textit{why}
testing is used.}
Employing socio-technical grounded theory (STGT),
we construct a theory by
systematically analyzing data from
19 in-depth, semi-structured interviews with software developers.
Allowing interviewees to reflect on
how and why they approach software testing,
we explore perspectives
that are rooted in their contextual experiences.
We develop eleven categories of circumstances
that act as conditions for the application and adaptation of testing practices
and introduce three concepts
that we then use to present
a theory of emerging testing strategies (ETS)
that explains why developers do (not)
use testing practices.
This study reveals
a new perspective on the connection
between testing artifacts
and collective reflection
of practitioners, and it embraces
testing as an experience
in which human- and social aspects are entangled
with organizational and technical circumstances.%
\end{abstract}

\section{Introduction}
For many decades, software testing has been considered a key component
of the software development process~\citep{hetzel_complete_1988}.
Systematic testing of software,
for example by using
unit tests is often practiced by software developers
to ensure a system's functionality~\citep{beller_when_2015,beller_developer_2019,Runeson_survey_2006}.
Because of its potential to prevent harmful software bugs~\citep{carstensen_lets_1995},
an urgency to better understand
the process of software testing
was signalled already in 2007
with a call to action~\citep{bertolino_software_2007}.
The field of software testing has been evolving for 40 years~\citep{gurcan_evolution_2022}
the landscape of approaches and tools getting ever more comprehensive,
guidelines for testing getting proposed~\citep{garousi_when_2016} and
skills needed to excel at testing being synthesised~\citep{sanchez_gordon_beyond_2020}.
\revised{%
Accordingly, opinions and
perspectives on testing
are manifold.
Students believe that testing is
monotonous and repetitive~\citep{santos_myths_2023}
and
work from~\citet{masood_like_2022} and~\citet{daka_survey_2014} 
has shown that testing is seen
as an undesired activity by developers.
However, \citet{santos_myths_2023} identify that testing
is also perceived as dynamic and exciting
and in a prior study we found that developers who write
sentimental posts on \stackoverflow{}
not only reveal negative views
but also positive, even aspirational attitudes
towards testing~\citep{swillus_sentiment_2023}.
Perceptions of testing range from
excitement and satisfaction to anxiety and stress
and challenges go beyond technicalities~\citet{hardman_testing_2025}.
\revisedtwo{By investigating the
rationalities behind those perceptions
through human-centered research,
sustainable improvements can be researched, devised and introduced to practice~\citep{sharp_role_2016}.}
The many facets of software testing practices
and the emotional connection between
people and their tools have recently also been put into the spotlight
by \citet{evans_scared_2021}.
They emphasize the significance of what they call \textit{testers' lived experience} (TX)
and propose it as a new area of research.
Indeed, since the call to action of \citet{bertolino_software_2007},
studies have repeatedly identified that
factors that go beyond technicalities
have an impact
on testing practices~\citep{martin_good_2007, rooksby_testing_2009, garousi_survey_2013, wiklund_impediments_2017}.
But despite the acknowledgement of their importance
no study has explored those factors in depth.
\revisedtwo{The rationalities behind developers' perceptions of testing
remain largely unexplored.
According to \citet{ardic_qualitative_2025},
technical- and practitioner-centered
research is still dominating the field of software engineering research;
few publication investigate the interplay
between human factors and more complex technical challenges.}
\revisedtwo{In this work we therefore
embrace the theme of TX
to explore when and why software developers do (not) use testing practices
and how their understanding, gained through first-hand experiences changes.}}{3.1}

\begin{itemize}
    \item[\textbf{RQ1}] What makes developers change their opinion about software testing?
    \item[\textbf{RQ2}] \revisedtwo{When do developers (not) use testing practices?}
    \item[\textbf{RQ3}] \revisedtwo{Why do developers (not) use testing practices?}
\end{itemize}

To understand
the factors that contribute
to developers' opinion about and use of testing,
we use socio-technical
grounded theory (STGT)~\citep{hoda_qualitative_2024},
a qualitative approach suitable for exploratory studies.
In this work \textit{opinion} refers to
the understanding of developers gained through lived experience.
We investigate the lived experience of developers, and
inquire when, and why developers are (not) using testing
practices.
\revised{%
We do not limit our inquiries
to practices like test-driven development (TDD)~\cite{shull_what_2010},
or techniques
like unit-testing, and
instead consider all systematic approaches to software testing.}{3.4}
We derive answers to our research questions
by comparing the perspectives that 19 developers have
shared with us through semi-structured interviews.
\revised{%
We choose an exploratory ground-up approach instead of
building on established theoretical theories like
the Technology Acceptance Model (TAM)~\citep{venkatesh_technology_2008},
Theory of Planned Behavior (TPB)~\citep{ajzen_theory_1991}
or Social Cognitive Theory (SCT)~\citep{bandura_social_1986}
because pertinent external variables
have not yet been identified
for the case of software testing.
Validation of generic variables and
exploration of practice specific variables is
recommended to avoid inaccurate interpretations~\citep{borstler_acceptance_2024, seuwou_user_2016}.}{2.1}

\revisedtwo{Work of~\citet{garousi_survey_2013}
and~\citet{martin_good_2007} 
shows that notions of testing rigor
are defined organizationally,
which is why we recruit participants with
varying organizational and cultural backgrounds
not only from our extended network
but also via the Q\&A platform \stackoverflow{}.
Rather than intricately describing one case
or the notions within a specific community of developers,
we aim to gather perspectives from a broad audience.}

\revisedtwo{By systematically comparing perspectives of 19 developers
we find that
the reasons for developers (not) to use testing practices
are rooted in the interplay of
organizational, technical and social
testing conditions.}
Technical solutions establish a foundation for developers
to contribute to testing efforts in an efficient way,
organizational aspects enable developers to allocate resources
for testing related activities, and social aspects
influence and encourage developers as they establish the
perceived value of testing for a project collaboratively (\textbf{RQ2}).
The collaborative reflection of testing experiences in
their unique organizational and socio-technical environment,
gives rise to developers' opinions about testing (\textbf{RQ1}).
Further, our analysis shows that
testing strategies are not placed or deliberately planned
by developers.
Instead, various conditions stimulate
a stochastic process that leads to
continuous adaption of testing strategies (\textbf{RQ3}).
Choices made by developers do not determine
but stimulate a recursive process from which
a testing strategy emerges. 
Social interactions
contribute to a testing culture and
technical contributions to testing infrastructure,
which together affect this process.
\revisedtwo{In order to understand
\textit{why} developers' choose (not) to use testing practices,}
one needs to investigate and understand how
both social and technical aspects
shape the conditions of testing
in the individual case.
In other words,
how testing experiences are embedded
in the organizational, technical and social
reality of a developer
and not just the testing experience itself
leads to testing decisions.

\revisedtwo{This paper makes the following novel contributions
which inform future research and can be applied when improving testing related processes and methods:
\begin{itemize}
    \item We present a categorized catalogue of
    organizational, technical, and social
    conditions that
    influence software developers' testing practices.
    \item We propose a theory of
    emerging testing strategies (ETS)
    that explains how those conditions
    affect developers' choices,
    which recursively lead to the emergence of
    testing strategies.
    \item We introduce three testing related
    concepts:
    \textit{testing signatures},
    \textit{testing echoes} and
    \textit{testing efficacy}
    to support the theoretical findings of our work.
    \item We situate our theory in the context of
    related work from
    software testing research,
    organizational aspects of software engineering,
    theories for acceptance of technologies,
    and organizational learning.
\end{itemize}}

The remainder of this paper is structured as follows:
First, we provide background information about
our research approach.
In \Cref{section:background} we introduce grounded theory (GT) and
then explain in \Cref{sec:method} how we employed it specifically for our
investigation by using the socio-technical version of GT (STGT)~\citep{hoda_qualitative_2024}
\revised{%
In \Cref{sec:theory},
we present the
results of our investigation
and propose our theory of emerging testing strategies (ETS).
Here, we first answer RQ1 by presenting
common narratives of interviewees on how testing
affected their careers.
We then develop three hypotheses
in \Cref{section:theory:emergence}
which underlie the theoretical
considerations of what follows in the remaining sections.
Additionally,
we introduce three theoretical
concepts in \Cref{section:theory:concepts},
and use them to explain,
what drives
software testing efforts.
We then use those concepts to theorize
how testing strategies evolve \Cref{section:theory:recursive}
and finally answer RQ2 and RQ3
in \Cref{section:theory:summary}.
To make the groundedness
of our considerations
transparent
we provide references to
interview excerpts (\Cref{appendix:interviews})
throughout the following sections.
Additionally, we present
further conceptual findings
of our work in \Cref{sec:conditions},
which resulted from the categorization of
interview transcripts.
The categorization of interviews
allowed us to
systematically compare and scrutinize the dataset.}{1.13}
In \Cref{sec:related_work}
we discuss our work.
We also specifically discuss related work in
\Cref{sec:form}.
Finally, we critically reflect our findings,
the research process and ethical implications in \Cref{sec:reflections},
before we conclude our work in \Cref{sec:conclusion}.
\section{Background - Grounded Theory (GT)}
\label{section:background}

\citet{glaser_discovery_2010}
developed Grounded Theory (GT) as an approach for qualitative research in the 1960s.
Ever since it has been used in many fields
by scholars with varying backgrounds.
The framework of GT
is made up of
data-gathering techniques,
strategies to analyze data
and guidelines
that help to
develop novel concepts and theories
in an iterative way.
In contrast to other qualitative approaches,
data collection and analysis are iterative and
interleaved \citep[$\S2.3\P5$%
\footnote{Instead of the page number
we provide a chapter indicator ($\S$) and
paragraph number ($\P$)
when we refer to or quote from
extensive publications like books
which are often re-published
in different layouts,
which can make page numbers
ambiguous.}]{hoda_qualitative_2024}
to sustain a high level of involvement
with the data~\citep[$\S5.1.3\P1$]{charmaz_constructing_2014}.
What also differentiates GT from other approaches
is its focus on understanding a given phenomenon
without being unduly influenced by existing concepts and theory.
The researcher starts with an open mind,
avoiding preconceptions until original concepts and theory
emerge from the data through rigorous analysis.

GT is agnostic about the kind of data that is analysed.
Depending on the research question,
an appropriate means of data collection
is chosen by the researcher.
Data collection strategies can also be combined.
For example, non-interactive documents like archive material
can complement interactive data gathering
through means of
interviews or participant observation.
Independent of the kind of data that is collected,
its analysis is done my means of
systematic coding techniques (e.g., open coding)
and constant comparison.
Through constant comparison of data and codes,
the researcher construct categories and establishes links
between them.
New insights that arise from this process
steer consecutive rounds of data collection,
until the researcher is able to
construct a theory that explains
the link between all categories.

The usage of GT by
scholars with various backgrounds,
resulted in the re-interpretation
of its original methods,
leading to the development of many different flavours of GT.
Crucially, approaches rest on different
epistemological foundations.
For example, the original Glaserian GT takes a
rather objective, positivist stance,
but Constructivist GT, which was proposed by Kathy Charmaz
moves away from positivism,
incorporating the beliefs and preconceptions
of the researcher into analysis~\citep[$\S1.3\P6$]{charmaz_constructing_2014}.
Consequently, flavours of GT differ in details
on how techniques for coding,
data gathering and theory development are executed
and how tightly strategies need to be followed.
For example, aligning with the constructivist view
Charmaz' version of semi-structured interviews
which we employ in our work embraces the idea of
co-creation of knowledge and experience.
In contrast, GT by \citet{corbin_basics_2008},
aligns with a positivist view and encourages
structured data collection,
including structured interviews.

Situating the GT approach into the field
of software engineering research, Socio-Technical GT (STGT) was introduced
to guide and ease application of GT in socio-technical fields,
\textit{where social and technical aspects are inherently interwoven}~\citep{hoda_socio_technical_2022}.
The STGT method is being applied to generate rich descriptive findings
and theories in software engineering~\citep{gama_socio_technical_2025},
artificial intelligence~\citep{pant_ethics_2024},
human robot interaction~\citep{chan_understanding_2023},
digital health~\citep{wang_adaptive_2024}
and other socio-technical (ST) disciplines.
It is particularly suitable for our study
as we investigate a ST phenomenon
(human experiences of software testing)
in a ST domain (software engineering),
studying ST actors (software developers),
and are ourselves ST researchers
(with backgrounds in software engineering
practice and research and theory development)~\citep[$\S3.1.2\P5$]{hoda_qualitative_2024}.
\section{Research Method}\label{sec:method}
\begin{figure*}[] 
	\centering
        \includegraphics[width=\textwidth]{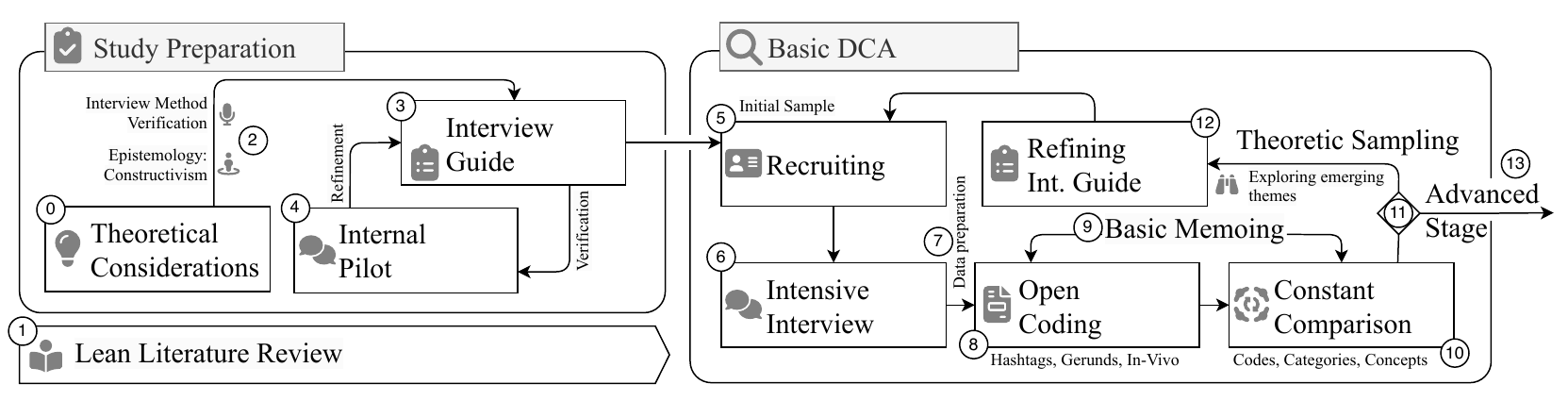}{}
        \caption{Illustration of the first phases of our STGT research design
        \citep{hoda_qualitative_2024}.
        We structure our research design in four phases.
        The \textit{Study Preparation}- and \textit{Basic data collection and analysis (DCA)}
        phases are illustrated here.
        The consecutive phases of \textit{Advanced DCA} and \textit{Theoretical Structuring and Reporting}
        are illustrated in \Cref{fig:advanced_stage}.
        Circled numbers~\circled{0} to~\circled{13}
        are referenced in the text.
        }\label{fig:basic_stage}
\end{figure*}

STGT is structured in two key stages:
Its \textit{Basic Stage} focuses on data collection and analysis,
its \textit{Advanced Stage} on theory development.
In this section we present how we approach
the two stages and the
steps and procedures lead from study preparation to theory reporting.
We use circled numbers (\circled{0} to \circled{24})
to refer to \cref{fig:basic_stage} and \cref{fig:advanced_stage}
which illustrate our research design.

\subsection{Basic Stage}

\subsubsection{Study Preparation}

Inconsistency in the application of research strategies
and inconsistency in the ontological and epistemological
perspective taken in research can threaten credibility
and applicability of research approaches~\citep[$\S5.1\P2$]{hoda_qualitative_2024}.
Before we begin evaluating and testing
data collection methods,
we therefore
reflect and explicitly declare
the philosophical stance of our research~\circled{0}
and conduct a lean literature review~\circled{1}
to scan the research area and to scope our work,
familiarize ourselves with applicable research instruments
and to guide our research design~\citep[$\S6.2.1$]{hoda_qualitative_2024}.

\paragraph{Epistemology and theoretical considerations}
Our stance with regard to our
research questions is that the reality of
testing practices and the experience of
practitioners in a complex environment is always unique.
An experience of one individual can never reflect
testing experience in its entirety.
Within the framework of STGT we therefore
take a constructivist stance~\citep[$\S5.5.1$]{hoda_qualitative_2024}
\revised{and selectively
employ techniques and considerations
of constructivist GT~\citep{charmaz_constructing_2014}
where applicable and in-line with STGT.}{3.6}
With our work we do not aim to find an objective truth;
instead we aim to describe what is common to and true for
various observers of the same phenomenon.
The theory we propose is therefore
not an objective representation
but aims to explain
phenomena which influence developers' subjective experiences.

\paragraph{Data collection method}
Through the analysis of interviews
we want to get insights
into the lived experience
of software developers
in the context of software testing.
We therefore choose
not to consider perspectives of
practitioners who are only carrying out testing tasks (e.g., QA-Engineers)
in this study.
\revised{We investigate experiences
of software engineers
around tasks that
have the objective to engineer production code.
We also aim to interview developers
who have substantial experience in
contributing to collaborative software projects.}{1.3}
Taking a constructivist stance
we acknowledge that
asking software developers about their lived experience
in interviews cannot produce
objective accounts of what their practice entails.
What developers tell us is always an incomplete
representation of experience
that is bound to their individual sensibility.
Developers might not be able to verbalize their intuitive understanding
of practice and how it relates to their experience;
they might not be conscious of influences
that lead to
their choices and opinions.
\revised{According to these considerations
we choose semi-structured interviews,
as suggested by \citet[$\S3.1\P1$]{charmaz_constructing_2014}
and recommended for STGT~\citep[$\S$8.3]{hoda_qualitative_2024},
as our data collection method~\circled{2}.}{3.6}
Semi-structured interviews
allow the interviewer to follow
unanticipated areas of inquiry,
hints and implicit views and accounts of action.
Instead of following a strict interview guide,
the researcher inquires about a topic through open questions
that gently guide
conversations with interviewees.
Instead of asking interviewees directly about specific aspects of their experience
gentle guidance aims at creating a space that allows
pertinent questions to arise.
We take conversation to deeper levels
by nudging toward specifics.
This allows us to go into depth. 

\paragraph{Interview guides}

We prepare an interview guide with
open questions and follow-up questions
for each round of interviews~\citep[$\S8.3$]{hoda_qualitative_2024}.
To systematically construct an interview guide for our first interviews
we follow recommendations by~\citet{kallio_systematic_2016}~\circled{3}.
We first construct a preliminary interview guide
with questions that direct conversation towards the research topic of software testing.
We formulate
questions which are participant-oriented,
not leading,
clearly worded,
single-faceted and open-ended
in order to provoke spontaneous,
in-depth and vivid answers.
\citet{kallio_systematic_2016} recommend
using \textit{what-, who-, where-, when-} and \textit{how-}questions
to achieve this.
We also add follow-up questions
which can guide participants
in case they find it difficult
to answer a question right away.
\revised{Additionally to an evaluation of
  interview guides following recommendations from STGT\cite[$\S8.3.3$]{hoda_qualitative_2024}
  we use Charmaz' extensive list of reflective questions
~\citep[$\S3.2.2\P10$]{charmaz_constructing_2014}
to fine tune our questions.}{3.6}
\revised{Finally, we conduct an internal pilot interview
to test the conditions under which we conducted interviews (e.g., duration, equipment)
and to assess the appropriateness and clarity of questions~\circled{4}.
We and adapt questions where needed.
We decided to include this initial interview in the final dataset
as the interviewee identifies as a software developer
in a university and research setting.}{3.7}
Aligning with the data collection approach of STGT (theoretic sampling)
we keep on reflecting on the need to adapt our interview guide
even after the pilot test.
We published all interview guides in the supplementary materials
of this work~\citep{swillus_interview_2026}.

\subsubsection{Data collection} 

Instead of following a sequential approach,
STGT follows iterative steps of data collection and analysis.
Data collection is immediately followed up by data analysis
and after each iteration the richness of data and
potential gaps are identified to motivate the next
round of data collection and analysis~\citep[$\S7.1\P2$]{hoda_qualitative_2024}.

\paragraph{Initial Sample}
We start the iterative process
of data collection and analysis
by conducting two semi-structured
interviews with
software developers who we consider
to be experts in the field of testing~\circled{5}.
After collecting and analyzing this initial sample,
we continue with the second iteration,
interviewing
three developers with less experience in testing
and we refine the interview guide accordingly~\circled{12}.

\paragraph{Theoretic sampling}
In GT, the purposeful and iterative recruitment of interviewees,
on the basis of emerging concepts and theoretical hunches
is called theoretic sampling.
Iterative rounds of data collection is done in
response to the emergence of theoretical concepts
or the identification of gaps~\citep[$\S2.3\P5$]{hoda_qualitative_2024}.
\revised{We therefore select interviewees carefully
with a goal to reach \textit{theoretical saturation}~\citep[$\S2.4$ purposive sampling and non-statistical generalization]{baltes_sampling_2022}.}{3.8}
We recruit participants
through the extended network
of our institution,
\stackoverflow{} posts,
and casual encounters.
\revised{In all cases we purposefully
recruit interviewees because sites,
specific presumed experience,
or their role (e.g., senior developer, manager)
fit the theoretical direction of our analysis.}{3.8}
Interview guides were adapted
accordingly for each round.
In \Cref{fig:basic_stage} we illustrate that the
data collection and analysis process is circular
until the maturity of analysis allows the researcher
to continue with the advanced stage of STGT~\circled{11}.

\paragraph{Interviewee Demographics}

\begin{table*}[]
    \centering
\setlength{\tabcolsep}{2.5pt}
\begin{smaller}
\begin{tabular}{c|p{1.8cm}|lllllll} \toprule
     Stage & Focus & Identifier & In- & Duration & Role & Exp. & Industry \\
     & & & person &  &  &  &  &  \\ \hhline{---------}
        & Pilot & \subj{Caro}{0}{0.Ac.E} & yes & 0:52:13 & Early & 1-5 &  Academia  \\\hhline{~--------}
       & Initial Sample  & \subj{Tim}{2}{1.Fi.X} & yes & 0:57:13 & Expert & 10+ & Finance  \\
    &  & \subj{Mauricio}{4}{2.Fi.X} & yes & 0:48:21 & Expert & 10+ & Finance \\\hhline{~--------}
    &   Less                                   & \subj{Niels}{6}{3.Fi.E} & yes & 1:03:09 & Early & 1-5 & Finance \\
    & Experience  & \subj{Julia}{8}{4.Fi.E} & yes & 1:07:35 & Early & 1-5 & Finance \\
    &    & \subj{Marlow}{10}{5.Fi.E} & yes & 1:05:41 & Early & 1-5 & Finance \\\hhline{~--------}
    &   Broaden  & \subj{Dan}{1}{6.Sw.S} & no & 0:25:08 & Senior & 10+ & Software \\
    &    Perspective  & \subj{Salim}{3}{7.Fi.S} & no & 0:35:02 & Senior & 10+ & Finance \\
    &                           & \subj{David}{5}{8.Sw.S} & no & 0:34:53 & Senior & 10+ & Software \\
    \multirow{-10}{*}{\rotatebox{90}{Basic}} &    & \subj{Csnad}{7}{9.Sw.S} & no & 0:32:45 & Senior & 5-10 & Software \\\hhline{---------}
    &                       Organizational                               & \subj{Marcel}{43}{10.Sw.M} & no & 0:42:13 & Manager & 10+ & Software  \\
     &  Aspects                                               & \subj{Kevin}{45}{11.Sw.M} & no & 0:50:15 & Manager & 10+ & Software  \\
     &   & \subj{Robert}{47}{12.Sw.M} & no & 0:48:17 & Manager & 10+ & Software  \\\hhline{~--------}
   & Theoretical & \subj{Maarten}{49}{13.Tr.S} & no & 1:24:13 & Manager & 10+ & Transport  \\
     & saturation & \subj{Arie}{51}{14.It.S} & yes & 0:50:44 & Senior & 10+ & IT Services  \\
    &   & \subj{Vincent}{53}{15.Re.S} & no & 0:40:21 & Senior & 5-10 & Retail  \\\hhline{~--------}
    &   Refinement                                                  & \subj{Simon}{55}{16.Sw.S} & no & 0:47:22 & Senior & 10+ & Software  \\
     &   and                                & \subj{Henning}{57}{17.Tr.S} & no & 1:18:34 & Senior & 10+ & Transport  \\
   \multirow{-9}{*}{\rotatebox{90}{Advanced}}&  verification          & \subj{Antioine}{58}{18.It.S} & no & 0:43:13 & Senior & 10+ & IT Services \\ \bottomrule
\end{tabular}
\end{smaller}
\caption{Participants in the order in which they were interviewed.
  The interview ID for each participant refers to the Industry
  they work in and the Role they carry out at the time
  of the interview. For example Interview \ref{subj:53} is interview number 15 with someone working in retail as a senior developer.}
\label{table:participants}
\end{table*}

In \Cref{table:participants} we provide an overview
of all 19 interviews we conducted.
Horizontal lines
indicate separate rounds of theoretical sampling
and data analysis.
We recruit the majority of participants
by reaching out to
individuals
through the extended network
of the researchers.
\revised{During the third iteration of data collection in the basic stage,
we recruited participants
from \stackoverflow{}.
We sent personalized invitations to
around 100 \stackoverflow{} users who had posted
questions about testing.
We selected posts by filtering the
public dataset of posts\footnote{\weblink{https://data.stackexchange.com/stackoverflow/queries}{data.stackexchange.com/stackoverflow/queries}}
using a list of testing related tags from our
previous work~\citep{swillus_sentiment_2023}
and then checking manually whether authors of posts
provided an email-address via their \stackoverflow{} profile.}{3.8}
We share limited demographic background of participants to
avoid revealing their identity.
\revisedtwo{Gender and age were not explicitly reported by participants.
However, one participant presented themself as female
and the rest presented themselves as male.
Their age is widely distributed between 24 and 50 years.
In \Cref{table:participants} we indicate
the industry\footnotemark{} in which participants work,
how long they have been working as developers,
and their current role using the following categories:}
\begin{itemize}
    \item \textbf{Senior Developers.} Developers with extensive experience in software development
        with no explicit focus on software testing during their careers.
    \item \textbf{Experts.} Senior developers who dedicate their work to testing and have extensive experience doing that.
    \item \textbf{Early Career Developers.} Developers who are in an early stage of their career, still being guided by mentors or trainee programs but nonetheless actively engaged in software development projects.
    \item \textbf{Managers}. Software developers who carry the responsibility to manage software projects and the people involved in them.
\end{itemize}

For each interview we planned a duration of one hour.
As \Cref{table:participants} indicates,
some interviews are significantly shorter or longer.
Shorter interviews are due to time constraints of our interviewees which we
incorporated by shortening the interview guide.
Longer interviews are due to lively discussions which were continued with
the explicit consent of interviewees.
Our own preference to do the interviews in person was 
met by six participants.
The remaining interviews were conducted
using online video communication platforms.

\paragraph{Conducting Interviews}
\revised{All interviews were conducted by the main author
and during all interviews
no one else apart from interviewer and interviewee was present.
As indicated in \Cref{table:participants},
7 of 19 interviews were done in person,
the rest using online video-chat platforms.}{1.6}
To facilitate the open approach
of semi-structured interviews \circled{6} and to
avoid a strong framing of narratives,
we conceal, where possible,
the topic of our research.
We always begin interviews
with a very general and open prompt:
\textit{Tell me something about you. Tell me about your experiences as a software developer}.
In most cases interviewees answered by
elaborating on how their interest in software engineering
was sparked and how their careers unfolded.
We use nudges to allow interviewees to explicate
their understanding and interpretation of
abstract terms or concepts to avoid imposing
our own preconceptions:
\intervieweenoref{The problem is that at university
you don't see the big
projects that you run at companies.
\interviewer{When you say big,
what exactly do you mean by
big software projects?}}
During STGT's basic stage
of data collection and analysis
we approach the topic of software development
more generally
until the topic of
software testing or software quality is mentioned by the
interviewee.
We do this to avoid creating a frame
that is too narrow for interviewees to
be able to reflect on the effect of testing
on their overall software development experience.
Once we are on topic,
we keep the conversation on topic,
by asking the interviewee
to elaborate on details
or by provoking reflection on testing experiences.
For example, by bringing up controversial ideas:
\interviewer{Someone once told me that
source code is never complete without tests. What do you think of that?}
We take such prompts,
open questions and follow-up questions
from the interview guide. 
At the end of the interview, we
take about ten minutes to 
offer the interviewee to give feedback
and ask questions.
To be able to fully concentrate on the conversations,
we record interviews
which we then transcribe.
We do write supplementary notes after each interview,
to record subtle details which are not audible,
like a visible affect when a specific
topic was brought up.

\paragraph{Data Preparation and Filtering}
To immerse ourselves in the
collected data
we manually transcribe
interviews during STGT's basic stage.
We automate transcription during the
advanced stage to speed up the process.
After importing the transcripts into a CAQDA\footnote{CAQDA: Computer-assisted qualitative data analysis}-software
we color code the text to make it easier to navigate~\circled{7}.
We highlight
prompts and questions asked by the interviewer and
use different colors for \textit{noise},
demographic information,
off-topic parts of the conversation
and on-topic~\citep[$\S9.2$]{hoda_qualitative_2024} parts.
We omit parts during data analysis that have
nothing to do with the subject of our study, however
we do include off-topic sections
that do not concern software testing
but other software engineering related topics
, as they can provide
context. 

\subsubsection{Data Analysis}\label{section:data:analysis}

We use the open coding technique as recommended in STGT~\citep[$\S10.3$]{hoda_qualitative_2024}
to begin the iterative process of
data analysis~\circled{8}.
During STGT's basic stage
we start without any preliminary codes,
remaining open to all possible
theoretical directions.
\revised{We code interview transcripts
line-by-line using \textit{hashtag} codes
to capture the socio-technical context of their experience~\citep[$\S10.3.1$]{hoda_qualitative_2024}
(e.g., \code{definitionOfDone})
and \textit{In-Vivo} codes,
which are quotations of what the interviewee said,
in their
own words
(e.g., \intervieweenoref{Who cares about testing? I had other things in mind.}\groundedin{4}{1.002})}{3.6}

We write analytical memos \circled{9}
about emerging codes and
reoccurring themes
during coding but also
when comparing emerging codes, categories and concepts~\citep[$\S10.5$]{hoda_qualitative_2024}.
Memos are analytical descriptions
of hunches, ideas and observations
used to record reflections about the work as it progresses.
For example, the following memo was written after
we conducted the first two interviews

\begin{textbox}[\faIcon{file-signature} Memo: Negotiating testing]{MEMO}
  Interviewee 1 explains how developers sometimes need to negotiate
  testing with project managers (\code{negotiatingTestingApproaches}\groundedin{2}{0.75}).
  \textit{Do I really want to write a piece of code that is not testable?
  Can we maybe, or should we maybe invest the time to refactor this?}
  A lot of social interaction and
  skill is required for this process.
  They also mention that you discuss past events.
  For example, how well some tests worked.
  And then you look at it and you
  \textit{come together as engineers}(\code{encouragingReflection}) to fix it\groundedin{2}{4}.
  The coming together here could be key.
  Coming together to establish testing
  practice and to build knowledge around testing
  seems to be an essential part of their experience.
\end{textbox}

The notion of testing being a practice
that is negotiated
in the above memo
was developed further
as we conducted more interviews.
In the advanced stage of data analysis
we use those memos
to develop preliminary hypotheses.
Reviewing memos that are written
especially in the beginning of the basic stage of
STGT can reveal in the advanced stage
how concepts and categories
were developed and how
they are grounded in data that is analyzed.

By constantly identifying differences and commonalities of interviews
and by reassessing the significance of all codes
within the same interview and across interviews,
we condense our analytical work
to advance theoretical directions.
In grounded theory this inductive approach is called \textit{constant comparison}
and leads researchers from specific instances
toward general, more abstract patterns~\circled{10}.
Using constant comparison we raise codes
to the level of concepts and category where applicable.

\subsubsection{Progression of Theoretic Sampling}

\revised{For each iteration of interviews we
purposefully select interviewees
who allow us to explore specific theoretical directions
and we adapt the interview guide accordingly~\circled{12}.}{1.4}
\revised{%
After analyzing interviews
from two experts on testing in our first round
of interviews we identify that the first contact
with testing significantly shapes their opinion about it.
We therefore recruit developers who are only at the
start of their careers for the interviews next.
Perspectives of less experienced developers
made it more evident in our analysis
that testing practices are not just a tool
for development velocity and software quality.
We identified that
testing can serve social needs like
the need for safety and confidence of developers.
This made us aware of the impact of collaborative
processes and organizational constraints.
We explored those themes further in the next round
and concluded the basic stage of data collection and analysis
by extending the diversity of perspectives
through purposive recruiting
of \stackoverflow{} users.}{1.20}
By employing strategies for constant comparison
as described above,
categories and concepts start to emerge from the data.
Iteratively extending our data set
we develop more and more refined codes
and preliminary categories
which become increasingly analytical
as we go forward.
Establishing links between analytical categories
we explicate, deepen and substantiate our analysis.
We now consider those relations
to construct
preliminary hypothesis
which explain the broad phenomena that transpire through our analysis.
At this point we reach the end of the basic stage and
decide to proceed with the advanced stage for theory development~\circled{13}.

\subsection{Advanced Stage}
\begin{figure*}[] 
	\centering
        \includegraphics[width=\textwidth]{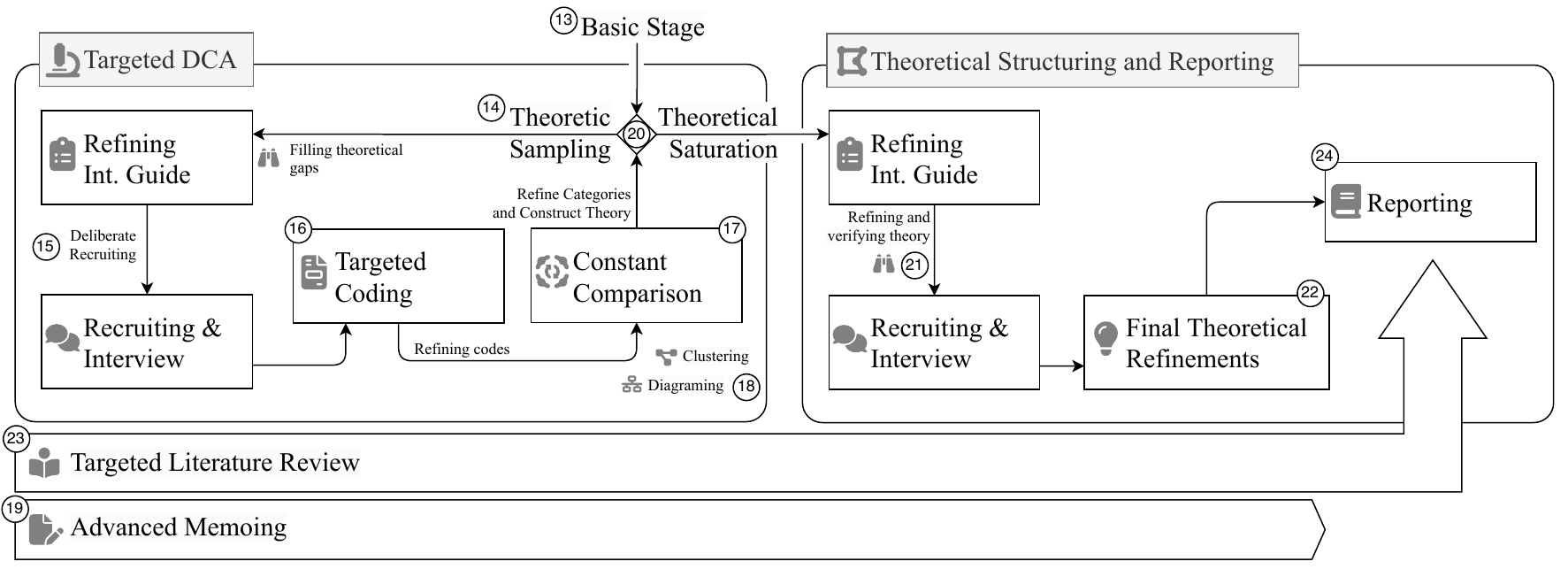}{}
        \caption{Illustration of the last two phases of our research design.
        We reach the advanced stage when data collection and analysis
        in the basic stage (refer to \Cref{fig:basic_stage}) allows us
        us to establish links between preliminary categories~\circled{13}.
        In the advanced stage for data collection and analysis we
        focus our work and begin the process of systematic theory construction.}\label{fig:advanced_stage}
\end{figure*}

The basic stage of our study
aims at exploring a broad phenomenon.
We construct categories and hypotheses
through data analysis in the basic stage,
but we only see indications
of relationships between them.
We lack evidence for
an overarching theoretical structure.
We therefore continue
in STGT's \textit{emergent mode}, which employs
a targeted strategy for theoretic sampling
and data analysis and theoretical structuring
as a means to construct theory~\circled{13}.
The emergent mode of theory development
allows for the theoretical structure
to emerge in an organic manner and
be finalized progressively~\citep[$\S12.5$]{hoda_qualitative_2024}.
We illustrate the advanced stage
for data collection and analysis
and theory development (\circled{14} to~\circled{24})
in \Cref{fig:advanced_stage}.

\subsubsection{Targeted Data Collection}
Proceeding with the advanced stage
we continue to use
theoretic sampling
as described above~\circled{14}.
Considering the categories
and hypothesis we developed in the basic stage,
we select interviewees that are most likely
to help fill theoretical gaps
and deepen our analysis~\circled{15}.
\revised{%
We begin the advanced stage of data collection
by interviewing three managers
who have been software developers
earlier in their career.
This choice is motivated
by the emergence of the following categories
during:
\code{TestingMandates},
\code{BusinessDomain},
\code{TestingCulture}.
Exploring the managerial perspective
allows us to integrate
and refine these categories.
Notably, we do not exclusively
focus or force those categories onto interviewees.
We still only gently guide interviews,
leaving enough room to explore
their perspectives in an open way.
Among other themes, the notion that
collaborative, undocumented reflection
can steer the process of testing strategy development
became more distinctive through the analysis of those interviews.
This motivated us to further investigate
the interplay of artefacts and collaborative reflection
in succeeding interviews with senior developers.}{1.20}

\subsubsection{Targeted Data Analysis}
We use
\textit{targeted coding}~\circled{16} to refine
the most significant codes, concepts
and categories from the basic stage.
We concentrate on those elements we have
already identified instead of remaining open to all possible
directions.
Unexpected findings are however
incorporated through the construction
of new categories where applicable.
Using constant comparison techniques we
strengthen links between categories~\circled{17}.
Aiming to formulate overarching explanations for phenomena
based on all available evidence,
we continue writing analytical memos
during the whole advanced stage of data collection and analysis~\circled{19}.
At this stage we focus memo writing
on categories, hypothesis and theory construction.
With memos we are also addressing gaps in our analysis
to track and inform the process of theoretic sampling
and to identify if and when theoretical saturation is reached.
Taking our research to the next phase of theory construction
we keep on writing analytical memos until
we begin writing our report.

Focusing more and more on theory construction,
we now also incorporate advanced
strategies for constant comparison.
We refine categories and their relations
using the \textit{operational model diagramming} technique
described by Salda{\~n}a~\citep[$\S5.4$]{saldana_coding_2013}~\circled{18}.
Through diagramming we explore
detailed features
of the coded dataset
from different angles.
For example,
starting with a code
like \code{encouragingReflection}
or a pertinent quote
that seems important but ambiguous
when categorized,
we sketch a network of connections
to other quotes,
categories or codes
on paper.
By making those relations
and the evidence
that supports it explicit we
take our analysis deeper without losing
touch with the data in which it is grounded.

We also use the \textit{clustering} technique
as described by
\citet[$\S7.2.1\P4$]{charmaz_constructing_2014}.
Grouping interview sections using categories
and writing analytical memos which
describe commonalities and differences
among those clusters we refine categories
and raise the level of our analysis.
Taking a different perspective each time,
we advance different explanations
for the meaning and value that testing
has to interviewees
and how it affects them.
We continue the process of analyzing the dataset using those
strategies
until we are able to construct a theory
that unites those different explanations~\circled{20}.
At this point, we reach theoretical saturation
where further collection and analysis
of interview data does not significantly
add to existing concepts or categories.
When interviews no longer yield new perspectives
we finalize our work.

\subsubsection{Theoretical Structuring}

\revised{Grounded theory studies are distinct
from other qualitative research frameworks
in their approach to build on already established knowledge.
Instead of adopting and building on an established
theory or theoretical framework,
for example Actor-Network-Theory~\citep{latour_reassembling_2007},
an STGT study starts,
as much as possible,
with a blank theoretical slate.
Especially in exploratory work like ours,
existing theory
is not integrated
before data collection and analysis
leads to theoretical results.}{2.1}
The advantage that this detachment
brings is arguably
also a weakness
which has been remarked
by scholars~\citep{giles_timing_2013, cutcliffe_methodological_2000}.
A researcher
is not an empty vessel
without biases
and approaching research
with an empty head
and without any sensibility
for known phenomena
risks producing results
which are detached and meaningless
to both practitioners
and other researchers.
\revised{%
To address this weakness we
review literature in a targeted way especially
after our findings are written down
and embed our theory in a larger context~\circled{23}~\citep[$\S6.2.2$]{hoda_qualitative_2024}.}{2.1}
We do this not only to situate our work
into the established body of knowledge
of our field.
We also connect our findings to work
that has not yet been identified as
applicable in our field.
Building bridges to other disciplines in this way
and presenting our
reflections on the work of others
in our publication we are offering
a new theoretical and conceptual
vantage point for others.
In \Cref{sec:form} we discuss
how our theory aligns with others
and explicate those vantage points.

\subsubsection{Constructing Interpretive Theory}

Synthesizing the insights
and hypothesis we obtain by engaging
with the data through the whole data analysis process described above,
we are systematically constructing an interpretive theory.
Interpretive theories aim to offer accounts
for what is happening,
how it arises
and explains why it happens~\citep[$\S9.1.2\P2$]{charmaz_constructing_2014}.
We utilize a
constructivist research paradigm
and incorporate pragmatist considerations:
We recognize that our statements can only correlate
our interpretation of the experience of individuals
with our own experience,
and the body of knowledge from the field that is
available and known to us~\citep[$\S\RN{1}.6\P1$]{mead_mind_1934}.
Further, aligned with our constructivist stance we recognize
that empirical observation is inherently \textit{subjective}.
Taking this stance we emphasize practice and action
rather than providing laws that ask for strict
falsifiability through \textit{objective} empiricism.
\revised{We do not offer a theory that predicts
  or provides instructions but rather explain when, why and how
  phenomena occur.
Such theories are also referred to as \textit{theories of understanding} \citep{gregor_nature_2006}.}{2.9}
\revised{Concretely,
the theoretical contributions of the present work
are constructed on the basis of what developers
recall about their lived experience.
We explain how their
knowledge and their assumptions about testing
might have been constructed,
and how developers seem
to act on their views.}{3.10}
By taking this approach of theory construction,
we want to make
the broad phenomenon of testing
in software development
and relationships between the two visible.
By proposing our theory we want to open up new vantage points
for our own and the future work of others.
We understand theorizing as an ongoing activity that will be continued
through future work.

\subsubsection{Refinement and Reporting}
While structuring and explicating
our theory
we conduct a final round of
interviews~\circled{21}
to refine and verify
our findings.
\revised{
  The analysis of
  those interviews
  did not reveal new properties
  or theoretical insights about
  our emerging theory.
  After incorporating
  the last interviews
  into our analysis \circled{22},
  we therefore
  start writing a draft to
  report our theory~\circled{24}.
}{3.12}
In \Cref{sec:theory} we report our
theory.
Conceptual findings present
analytical categories and how they are
grounded in the data
are reported in \Cref{sec:conditions}.
\section{Emerging testing strategies - A socio-technical grounded theory}\label{sec:theory}

This paper explores
the lived experiences
of software developers
to understand
why testing practices
are (not) used by developers.
Our investigation reveals
a complex system of technical-, social- and organizational
conditions of software testing to us.
\revised{%
In this section we contextualise testing practices through
the lens of developer experience in order to get a hold on this complexity.
In \Cref{section:results:anecdotes} we
recognize that experiences with testing
are not purely technical
and argue that understanding changes through
a reflective process that is rooted in
profound and sometimes even emotional experiences (RQ1).
Then, in \Cref{section:theory:emergence}
we present how the interplay of the various conditions
of software testing influences testing decisions of
developers.
We identify that
the complex interplay of conditions
rather than a simple summation of conditions
shapes testing strategies.
In \Cref{section:theory:concepts}
we introduce 3 novel theoretical concepts
in order  to explain
how the interplay of conditions
leads to the evolution of testing strategies (\Cref{section:theory:recursive}).
Testing strategies we argue are not the result of a linear
process but emerge from a recursive process.
Finally, in \Cref{section:theory:summary} we answer
RQ2 and RQ3,
explaining when developers test and
why testing practices
are (not) used by developers
in practice.}{1.9}

The theory of Emerging Testing Strategies (ETS)
we present in this section,
is grounded in the data we
gathered from interviews.
Quotations which exemplify
the groundedness of our theory
in the perspectives of interviewees
are referenced using superscript numbers in parentheses,
e.g., \groundedintwo{4}{2}{10}{1}.
All numbered interview excerpts can be found in \Cref{appendix:interviews}.

\subsection{Software Testing Experience and Understanding}\label{section:results:anecdotes}

\revised{%
Giving interviewees the opportunity to
reflect on their experiences
brought stories and anecdotes to interviews.
\reviewmarktwo{Interviewees told us that
remembering those stories was a valuable experience\groundedintwo{4}{2}{10}{1}}}{1.1}.
  Interviewees report mind-opening experiences\groundedin{5}{1},
  spectacular failures\groundedin{5}{0} and
  encounters with critical bugs\groundedin{58}{1},
which fundamentally changed their perspective on software development
and software testing.
Across those narratives we find several recurring themes.
\reviewmarktwo{Developers consistently describe that concrete events
trigger them to reflect\groundedin{5}{5}
on how they develop software\groundedintwo{4}{1}{4}{1.6}
with a desire to change the foundations of their approach.}
As unexpected failures are often the cause of those events,
software testing suggests itself as a remedy for future failures.
\revised{We argue that this link
between transformative moments
and testing practices indicates that
opinions (understanding gained through lived experience)
about testing
have a tendency to be rooted in emotional
or otherwise profound experiences.
Developers' reflections concern not only
their own efforts but the social and technical environment in which
the event unfolds\groundedin{5}{3}.}{1.21}
Realization of responsibilities\groundedin{58}{2}
and the collaborative effort to resolve issues during a post-mortem
contribute to the rearranging of technical capabilities
and structures that go beyond the code\groundedin{58}{3}.
Through books and online resources like blogs or
question and answer platforms like \stackoverflow{},
even the view of the broader (online) community is considered\groundedintwo{5}{1.5}{4}{1.5}.
Consistent in the narratives of our interviewees
is the notion that beyond
the creation of technical artifacts,
collaborative reflection of practices
through interaction between developers
leads to changed opinions about testing and
adaptation of testing strategies.
Through collective reflection,
a common sense of the value of software testing,
which we call a \textit{testing culture}
is constructed\groundedin{43}{1}.

How experiences can trigger a cascade of changes
through reflection is apparent in cases of \textit{spectacular failure}\groundedin{5}{0},
but also identifiable in less drastic experience\groundedin{43}{1}.
Testing experiences are commonly described as involving communication\groundedin{45}{0.5},
encouragement\groundedintwo{49}{1}{7}{1} and
constant re-prioritization \groundedin{51}{1}
due to non-technical factors.
We therefore hypothesise that a social element in testing experience
provides meaning to testing efforts and thereby shapes
developers' opinion and the development of testing strategies.
Central to this social element is the activity of ongoing collaborative reflection
and a consequential re-prioritization and adaption of technology and strategies.

\begin{questionbox}{What makes developers change their opinion about software testing?}{1}
  Testing experience is made up of transformative experiences
  which can feel emotional, profound or meaningful to
  software developers.
  Those transformative experiences
  trigger reflections about the way in which software development is practiced
  and contribute to an
  emotional connection between people and their tool sets.
  \revised{The socio-technical environment}{2.7}
  in which testing practices are experienced,
  including views to which developers are exposed by engaging
  with the broader software development community
  further shapes this connection.
  Opinions about testing are linked
  to this connection as well and are changed through
  experiences which feel emotional,
  profound and meaningful to software developers (a posteriori).
  We only find suggestive evidence
  that character traits or other predetermined factors
  like technicalities contribute
  to opinions (a priori).
  We therefore argue that
  reflecting on meaningful experiences
  (e.g., introducing critical bugs)
  in the context of a project's
  testing culture and
  (online) communities
  in which
  ideas and opinions are \textit{performed}
  is what makes a developer change
  their opinion about software testing practices.
\end{questionbox}

\subsection{Conditions for Testing Strategy Emergence}%
\label{section:theory:emergence}

  We find that
  developers' experience of software testing
  is influenced by
  a wide range of organizational, social and technical factors
  such as organizational mandates for testing,
  to which extent developers take on responsibility,
  or the availability of technical building blocks to achieve their testing goals.
  We present all conditions and their categorization in \Cref{sec:findings}.
  We argue that the effect of those factors on testing choices
  cannot be understood in isolation.
\revised{
  We argue that testing strategies
  and their execution
  are influenced by a complex combination
  of interdependent conditions.
  For example, interviewees describe the perceived complexity
  of software projects as a driving factor for testing.
  Complexity can be both motivating~\groundedin{10}{0.5}
  and de-motivating\groundedintwo{6}{4}{51}{3}.
  \reviewmark{Complexity can have a positive and negative effect on testing,
  because the effect of its potential
  depends on factors such as the availability
  of testing infrastructure\groundedin{10}{0.7},
  tools\groundedin{45}{1},
  and working examples\groundedin{2}{0.5}.
  \reviewmarktwo{Working examples and testing infrastructure
  impact how developers consider the
  potential of their investment into testing~\groundedintwo{2}{1}{4}{1.6}.}}
  Further, considerations of developers are not only
  impacted by technical factors.
  For example, the testing culture within an organization
  influences developers'
  acceptance~\groundedin{58}{1} or aversion\groundedin{51}{1}
  of testing.
  \reviewmark{Testing culture can declare
  parts of a project as untestable\groundedin{2}{2},
  despite a recognized need for testing\groundedin{51}{2}.}
  \reviewmarktwo{Further, dogmatic views\groundedin{4}{1.5},
  can even declare sensible practices to be crazy\groundedin{4}{1.2}.}
  Whether a particular situation warrants testing does not depend on the
  perceived complexity of a project but how this complexity can be dealt with
  in connection to other aspects which condition testing on a technical and non-technical level.}{1.8}

  \begin{hypothesisbox}
  \textbf{testing decisions} are affected by
  the \textbf{interdependence of conditions for testing}
  such as project complexity, testing infrastructure,
  and the testing culture of a project.
  \end{hypothesisbox}

  As developers make testing decisions
  and engage with testing,
  they change
  the technical factors (e.g., test suites)
  on which their decisions were based\groundedintwo{1}{1}{4}{1.01}.
 \revisedtwo{By communicating their changes,
  their decision also affects
  the social-, and organizational- factors
  which influence future decisions of peers (e.g., testing culture)\groundedintwo{53}{3}{10}{0.85}.
  The choice (not) to use testing practices.}
  alters the conditions that originally necessitated the decision and
  similar situations rarely unfold in the same way twice\groundedin{58}{1}.
\revised{Given this
  recursive development of conditions for testing,
  clear-cut causes
  are not suited to adequately explain what
  leads to choices in the context of the experiences
  interviewees shared with us.
  We argue that developers test
  when they \textit{see it fit}.
  This individual decision is
  influenced by conditions which impact
  the efficiency with which
  the developer can approach testing tasks
  and the value which they attribute to
  their testing effort
  in a particular, unique environment.

\begin{hypothesisbox}
    \revisedtwo{The decision (not) to use testing practices
    is constructed individually.}
    \revised{It is constructed case by case and
    bound to the concrete software development project.}{2.7}
    \revisedtwo{Causes
    for the
    decision (not) to use testing practices
    are not generalizable
    but can be identified through case by case
    investigations
    into specifics.}
\end{hypothesisbox}

  Accordingly, we argue that reconstructing
  how conditions influence
  a team of software developers
  to developed successful testing strategies
  is unlikely to yield
  universal guidelines for the construction of successful
  testing strategies\groundedin{5}{2}.}{1.1}

  \begin{hypothesisbox}
    Slight differences
    in the conditions for testing
    can lead to significant differences in
    the impact of testing approaches and decisions.
    \revised{
    Testing choices made in a specific project at a specific time
    are unlikely to yield the same result again
  when conditions are only slightly different.}{2.7}
  \end{hypothesisbox}

  \revised{
  Instead of proposing generalizable explanations
  (of the effect of conditions on testing adaption
  and adoption
  or the cause of successful or unsuccessful attempts),
  we approach research questions 2 and 3 on a more abstract level.
  In the remainder of \Cref{sec:theory}
  we first construct three concepts that
  allow us to distinct between identifiable
  aspects which contribute to the testing process.
  We then use those concepts to describe
  of how the evolution of testing strategies
  takes place.
  On the basis of those theoretical explanations
  we then answer \textbf{RQ2}: \revisedtwo{When do developers (not) use testing practices?}
  by identifying what stimulates individual testing decisions.
  Describing the nature of this process of testing evolution,
  we can then answer \textbf{RQ3}: \revisedtwo{\textit{Why} do developers
  (not) use testing practices?}}{1.1}

\subsection{Conceptual Foundations for a Theory of Emerging Testing Strategies}%
\label{section:theory:concepts}
\revised{We find that testing conditions
  affect developers in three ways.
  \reviewmark{First, developers experience testing through material
  (e.g.,
  documentation\groundedintwo{2}{0.15}{55}{2},
  infrastructure\groundedin{10}{0.7}).
  Second, testing is experienced through formal communication
  (e.g., code reviews\groundedin{49}{1}, online platforms\groundedin{5}{5})
  and informal conversation\groundedin{2}{3}.}
  \reviewmarktwo{Third, organizational conditions influence
  developers' autonomy and ability to engage
  in testing experiences
  (e.g., mandates\groundedin{3}{2}, available resources\groundedin{4}{1.2}).}
  To describe how these three dimensions
  stand in mutual relation,
  we introduce three novel concepts:}{1.1}

\begin{itemize}
    \item \textbf{\code{testingSignatures}} are testing related traces in
      technical, material artefacts
      (e.g., test-code, documentation, CI/CD-pipelines) that exemplify
      how testing is actually done.
      They signal the relevance of testing for a project
      to developers who interact with them\groundedin{47}{1}.
    \item \textbf{\code{TestingEchoes}} are short-lived verbal and non verbal
      impulses (e.g., conversation, blog-posts, discussions)
      that describe and carry
      perspectives on testing\groundedin{1}{3}.
    \item \textbf{\code{TestingEfficacy}} is the feeling of a developer
      of having the power to produce something desired with their testing efforts\groundedin{4}{1.002}.
      \reviewmark{It concerns ones own appreciation of
      efficiency and effective practices or fun\groundedin{1}{2}
      but also the value of testing for a project's users and collaborators\groundedin{2}{0.1}.}
\end{itemize}

\revised{%
  The concepts we propose consider
  three perspectives on testing.
  They enable us to theorize how testing strategies are constructed
  through experiences which are shaped by
  developers' communication (\code{testingEchoes}),
  their interaction with testing artefacts (\code{testingSignatures})
  and their perception of the value and applicability of testing (\code{testingEﬃcacy}).}{1.10}

\begin{figure}
	\centering
        \includegraphics[width=.80\textwidth]{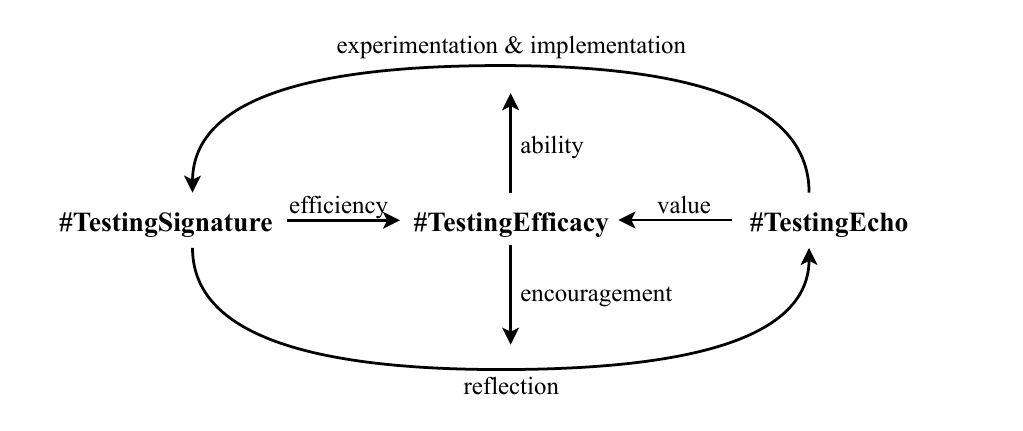}{}
        \caption{\revised{Diagram illustrating the mutual, circular influence of \code{TestingSignatures}
          and \code{TestingEchoes}.
          For example, a conversation between two developers (\code{testingEcho})
          about testing leads to experimentation
          and eventually to the creation of a new unit test (\code{testingSignature}).
          In turn, working with a unit test
          causes a developer
          to reflect.
          Those reflections about how and why the unit test does (not)
          fit into the project's testing strategy leads to new conversations
          in code reviews, meetings or informal settings.
          \code{testingEfficacy} influences this mutual influence.
          The presence of \code{testingSignatures} impacts the efficiency with which testing
          contributions can be made; existing unit tests can be copied.
          Conversations explicate the value of testing.
          In turn, the empowerment of developers to contribute
          to testing efforts in a way that feels
          efficient and appreciated (\code{testingEfficacy}).
          encourages developers to reflect
      and gives them the ability to experiment.}{1.10}\label{fig:minimal}}
    \end{figure}

\reviewmark{\code{TestingEchoes}
(e.g., discussions about a testing approach\groundedinfour{2}{0.75}{4}{1.2}{10}{0.85}{45}{0.5})
which resonate with project members,
produce new shared ideas about testing\groundedin{2}{3}}
and stimulate actions\groundedin{49}{1}.
They can lead to experimentation\groundedin{7}{1},
and to the implementation or adaption
of artefacts.
When artefacts are created or changed in this way,
developers inscribe
traces of those \code{testingEchoes}
in them\groundedintwo{55}{2}{47}{1}.
What was discussed is thereby partly represented in the artefact.
We call those traces or representations \code{testingSignatures}.
As software developers are in constant interaction
with artefacts,
\code{testingSignatures}, once inscribed,
act as symbols and exemplify
how testing can and should be done.
\reviewmark{Even many years after their creation,
when the original contributor might not be around anymore\groundedin{2}{0.15}.}
\code{TestingSignatures} exemplify in a technical language,
what is expected of
developers and what they
can expect of others\groundedin{43}{2}.
The presence of \code{testingSignatures}
stimulates developers to reflect.
\code{TestingSignatures} thereby carry
the potential for new \code{testingEchoes}
to be brought into conversations or discussions.
In \Cref{fig:minimal} we illustrate
this mutual connection of
\code{testingSignatures} with \code{tesingEchoes}.
The connection indicates a feedback loop:
Previous discussions (\code{testingEchoes})
influence how a newly voiced
\code{testingEcho} is received
and how likely it is that it leads to the creation of new \code{testingSignatures}.
Additionally,
the more
\code{testingSignatures} are
present,
the more likely it is
that reflections generate new \code{testingEchoes}.
Each new \textit{echo} and \textit{signature}
creates potential to bring testing strategies
into a motion.

\code{TestingEfficacy}, which we define as the
perceived ability or power
of developers to produce something valuable
by contributing to testing efforts
influences this circular dynamic.
The ability to engage in experimentation and implementation
of testing features\groundedin{4}{1.002} is supported by the
availability of resources like time\groundedin{55}{1} and tools\groundedin{5}{1.5}
to get the job done.
Whether \code{TestingEchoes} can be effectively
turned into \code{testingSignatures} therefore depends on
socio-technical and organizational factors that
facilitate \code{testingEfficacy}.
In \Cref{fig:minimal} we indicate the influence of \code{testingEfficacy}
on both the process of reflection and experimentation.
The connection of those processes and \code{testingEfficacy} is circular as well.
Experimentation can lead to the creation of \code{testingSignatures} in a project,
which change how efficiently new contributions and thereby new \code{testingSignatures} can be added.
\reviewmark{For example, re-using existing test code
makes the development of new test cases more efficient and approachable\groundedin{2}{1}.}
Or contrarily,
test suites can stifle
experimentation and progress when each small change triggers
an unmanageable amount of tests to fail\groundedin{1}{1}.
Whether \code{testingSignatures} spark new ideas and lead to \code{testingEchoes} is
also bound to socio-technical (e.g., trust\groundedin{43}{3}) and organizational factors\groundedin{53}{2} that facilitate \code{testingEfficacy}:
The social- and human needs of developers
and the way in which a project is organized impacts how developers
interact with and reflect about development.
\reviewmark{For example,
the code review process can be used to
reflect on how testing is done
and could be done differently\groundedintwo{49}{1}{2}{0.2}.}
Finally, collective reflection
through \code{testingEchoes} feeds back into
the \code{testingEfficacy} experienced by developers.
Developers participating in discussions
about testing practices
internalize how valuable specific contributions are
for the team.
Engaging in discussions, developers
collaboratively establish
the value of their testing contributions.

\subsection{Recursive Evolution of Testing Strategies}\label{section:theory:recursive}
As we illustrate in \Cref{fig:minimal},
organizational- social- and technical factors
influence each other in a circular way.
Altering organizational, technical or social factors which affect
either the potential for \code{testingEchoes},
the presence of \code{testingSignatures}
or the \code{testingEfficacy} of a project
brings the whole socio-technical processes around testing into motion,
leading to a cascade of changes.
The result of an attempt to change the testing strategy
therefore depends on the current
configuration of the whole system.

\begin{figure}
	\centering
        \includegraphics[height=0.3\textheight]{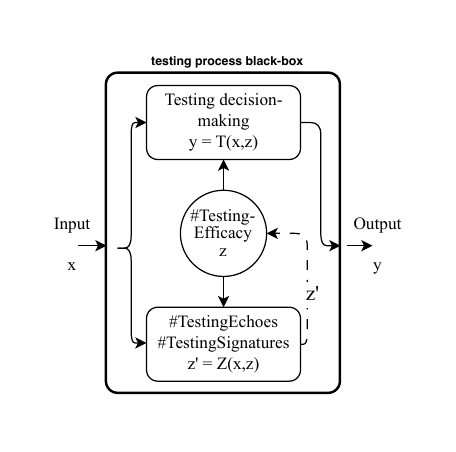}{}
        \caption{Depiction of the testing process as a black-box with recursive
          dependency $y=T(x,z)$. \code{TestingEfficacy} ($z$)
          influences the outcome of the testing decision making process $y$,
          and the potential of \code{TestingEchoes}
          and \code{testingSignatures} on efficacy for testing in the future ($z'$).
          As the actor has no full insight on $z$,
          their action cascade recursively
          and largely unpredictable \citep[see generally][pp.310,311 comparison of trivial and non-trivial machines]{von_foerster_understanding_2003}%
      \label{fig:non-trivial}}
\end{figure}

As we suggest in Hypothesis~\ref{hyp:3},
every contribution to testing may
change interdependent factors which influence the impact of
future contributions.
We illustrate this recursiveness in \Cref{fig:non-trivial}.
Every action or input to the testing process
(e.g., the task of implementing a feature)
leads to testing decisions made by developers,
which are conditioned by the \code{testingEfficacy} of a project.
Consider big, mature projects,
in which developers only have a partial view
of the configuration of factors that influence their efficacy to test.
From the perspective of developers,
the evolution of testing strategies
therefore resembles a stochastic and not a linear process.

\begin{figure}
	\centering
        \includegraphics[height=0.3\textheight]{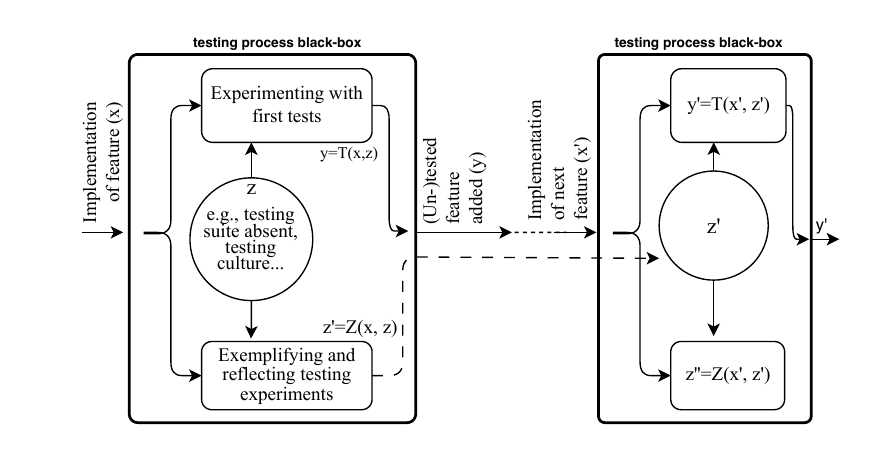}{}
        \caption{Illustration of an example of the recursive evolution of testing strategies.
          A developer implementing a feature unavoidably affects a projects testing strategy and
          future attempts at software testing. Each action $x$, like the implementation of a
          new feature leads to a testing related output $y$.
          The developers' decision to implement tests is affected by organizational,
          social and technical affordances $z$.
          Apart from the outcome $y$, the actions of the developer alter the conditions for testing ($Z\rightarrow z'$).
          For example,
          the developer presents the benefits of their solution for testing
          in a meeting which impacts how peers perceive testing (change in testing culture).
          The implementation of the next feature $x'$ is influenced by those altered factors ($y'=T(x',z')$
          with $z'=Z(x,z))$.
        \label{fig:non-trivial-example}}
\end{figure}

Interviewees describe
the recursiveness of testing processes using
the metaphor of an avalanche.
\reviewmarktwo{\reviewmark{To them, \textit{getting the snowball rolling}\groundedin{2}{1} means
establishing conditions to create
enough momentum,
so that the process we describe
steers testing efforts into the
desired direction\groundedin{4}{1.01}.}}
Efforts to improve testing infrastructure (\code{testingSignatures})
makes testing practices more approachable
to others (\code{testingEfficacy}).
Exemplifying the ease with which testing can now be pursued (\code{testingSignatures})
underpins arguments for a more rigorous testing practice (\code{testingEchoes})
which in turn motivates developer to engage in testing activities (\code{testingEfficacy}).
In \Cref{fig:non-trivial-example} we sketch
how this cascading effect can start to unfold.
When a feature is developed ($x$),
technical aspects, in part, determine how well and
how extensive the feature can be tested.
Beyond those technical parameters, the decision
of a developer to test is influenced by
organizational and socio-technical conditions.
For example, their prior experience with testing,
the presence or absence of tools,
the time available or
the testing culture within a software project~($z$).
Given those circumstances the developer chooses,
in the case we illustrate in \Cref{fig:non-trivial-example} to experiment
with testing, adding a minimal testing suite to the project that covers
the implemented feature.
Crucially, this decision that leads to testing contributions
($T\rightarrow y$)
is dependent on the development task ($x$)
and the sum of all technical and non-technical circumstances of the project~($y=S(x,z)$).
In the bottom half of the left box in \Cref{fig:non-trivial-example},
we illustrate
that the circumstances ($z$)
which lead to the decision
are also altered in the process ($z'=Z(x,z)$).
During experimentation the developer
discusses and demonstrates their testing experiments,
for example in code reviews or meetings.
By engaging with testing in this way,
not only the technical foundation for testing is altered
but also the testing culture in a team that might be more open
to investing more time into it.
Crucially, just like the decision to test,
these changes of the development environment
are themselves dependent on prevalent
socio-technical conditions~($z$).
\reviewmark{For example, when the team culture or development process
allows developers
to reflect on their works (e.g., in social settings\groundedin{2}{3}, through code reviews \groundedin{49}{1})
it is more likely that testing experiments
will engage collaborators.}
Finally, \Cref{fig:non-trivial-example}
illustrates that the testing process
is recursive ($y'=S(z',x')$ with $z'=Z(x,z)$).
When the next feature ($x'$) is developed,
conditions are altered.
The momentum that was
put into \textit{getting the snowball} rolling
introduced the potential for cascading changes.

\subsection{When and why developers test}\label{section:theory:summary}

This paper sets out to investigate when and why
developer chose to test and how their opinions about software testing change.
Exploring the developers' perspective we
learned that the conditions for decision making processes are
interconnected and form a complex system.
At the core of our theory (illustrated in \Cref{fig:minimal})
we identify \code{TestingEfficacy} as the prevalent
factor that contributes to
developers' ability to contribute to
testing efforts
and an encouragement to reflect
on how testing ought to be used.
Concretely, developers implement test-suites and experiment with testing technologies
when they experience the power
to produce something desired with their efforts.
This not only concerns the value that they attribute to those practices
in the confines of the project but also the efficiency with which
they can contribute with testing.
We summarize those findings in our answer to RQ2:

\begin{questionbox}{When do developers (not) use testing practices?}{2}
  \revisedtwo{Technical, organizational and social factors influence
  the conditions under which developers decide (not) to use testing practices.}
  When organizational-, social- and technical
  conditions in a software project
  support developers to efficiently pursue testing
  in a way that is perceived as valuable for the project,
  they engage in testing.
  When the present conditions in a project are such that
  testing is not perceived as a worthwhile activity
  or as too labour intensive to get started with,
  it is avoided.
  We call the entanglement of technical
  and non-technical conditions
  which contribute to the efficiency
  and perceived value of testing strategies in projects
  \code{testingEfficacy}.
\end{questionbox}

We argue that a dualism of testing artifacts on the one hand
and verbal communication and social interaction on the other
leads to the adaptation of testing strategies.
As developers contribute to test-suites
and other infrastructure
for the purpose of software testing,
they inscribe
how testing should be done.
We call those inscriptions \code{testingSignatures}.
Further,
\code{testingSignatures} shape technical conditions
of testing.
For example, the extension of a test-suite
can make testing more efficient.
Complementary, when developers talk about software development,
for example in meetings, during code-reviews or at the coffee table,
they might bring up software testing topics.
These impulses,
which we call \code{testingEchoes},
are not written down or otherwise documented.
Through \code{testingEchoes} developers establish
a shared value of testing.
Even though they are fleeting,
they motivate
experimentation with and
implementation of testing artifacts
and can have long lasting impact.
As illustrated in \Cref{fig:minimal}
and explained in \Cref{section:theory:recursive},
we theorize that the mutual effect of
the duality of \code{testingSignatures}
and \code{TestingEchoes}
drives a recursive process that
leads to the
emergence of testing strategies.
On the basis of this analysis we answer RQ3:

\begin{questionbox}{Why do developers (not) use testing practices?}{3}
  We argue that software testing strategies
  emerge from a recursive process
  which is influenced
  by conditions that are not always transparent to developers.
  Projects demonstrate consistent aversion or
  a long-lasting embrace of testing
  because of a convergence of this recursive process.
  When the recursive testing process converges,
  developers re-affirm
  established practices and views with their action.
  Testing contributions which
  feel valuable and approachable
  are embraced leading to a greater potential
  for future testing contributions.
  Testing contributions which
  feel worthless and too labour intensive are avoided
  making future testing contributions even more expensive.
  We identify that actions of individual developers
  impact this recursive process through
  what we call \code{testingEchoes} and \code{testingSignatures}.
  Communicating about testing practices (\code{testingEchoes})
  and exemplifying testing approaches,
  for example through code (\code{testingSignatures})
  adds momentum to the recursive process,
  potentially steering it into of a new direction.
  We argue that the reason why developers change their testing strategy
  and their testing decisions
  is an incremental change of the configuration of conditions
  for testing (\code{testingEfficacy}) through the cascading effect
  of those impulses.
\end{questionbox}
\section{Conceptual Outcomes -- Conditions of testing}\label{sec:findings}
In this section we present analytical categories
we constructed in the process of data collection and analysis.
Each category represents common themes
which touch on the experience of
software testing.
By comparing those themes with each other
and by interpreting how each theme
conditions testing practices,
we arrived at the theory that is presented in the preceding section.

For the remainder of this section
we present 12 analytical sub-categories
which are grouped in three main categories:
\begin{itemize}
    \item \textbf{Socio-technical aspects}. Conditions which influence technical and social complexity and how projects are organised.
    \item \textbf{Affordances}. Conditions which determine the technical and organizational framing of testing, including artifacts and business context
    \item \textbf{Dogmatic Perspectives}. Conditions which affect the ideas developers are exposed to
\end{itemize}

We describe each category
and present our interpretation of how
each of them
conditions adoption and adaption of
testing practices.
Quantifying the strength of the evidence
of our qualitative work and presenting it
as a proof for the validity of our arguments
would be misleading and
in conflict with our epistemological stance.
To indicate the strength of evidence
we instead provide a qualitative
indicator for each category signaling
how prevalent
the category was in
our interviews:

\begin{enumerate}
    \item \weakevidence{} (Weak/suggestive evidence): Evidence from multiple interviews support our arguments. However, the evidence is not robust indicating a possible association worth further investigations
    \item \evidence{} (Moderate/compelling evidence):
        Findings from multiple interviews are consistent. Evidence warrants consideration in future investigations and practice, but is not conclusive
    \item \strongevidence{} (Strong/concluding evidence): Evidence from a majority of interviews. Findings are consistent across interviews and can be considered conclusive
\end{enumerate}

\label{sec:conditions}

\subsection{Socio-technical Aspects}\label{sec:socio-technical}

\begin{figure*}
	\centering
        \includegraphics[width=\textwidth]{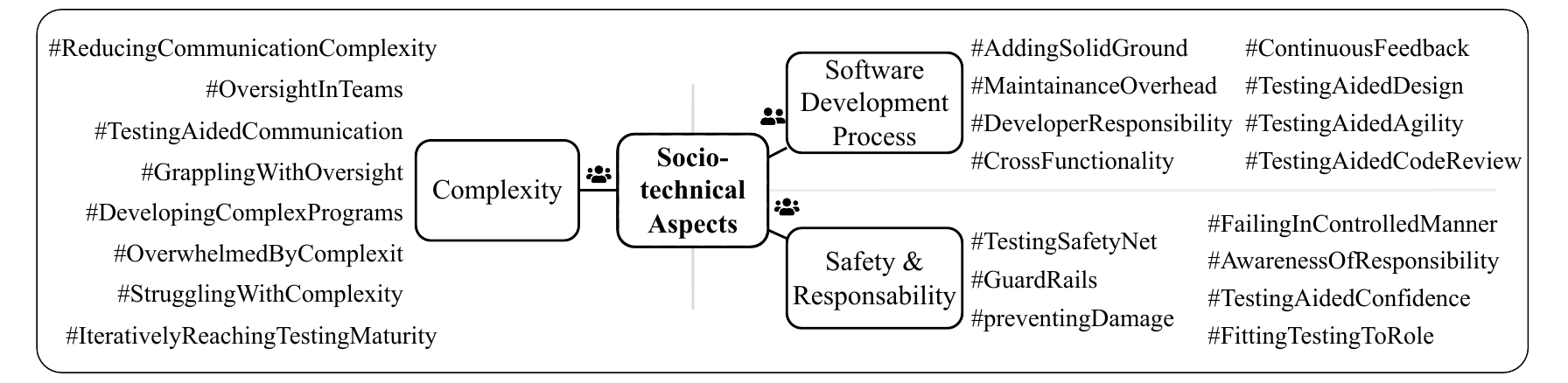}{}
        \caption{Diagram showing subcategories and
            the most prevalent codes which contributed to the forming of
            the analytical category \textit{socio-technical aspects}.
            Through constant comparison of interviews through codes we identified
            conditions that influence when developers use testing practices.}\label{fig:socio-technical}
\end{figure*}

We categorize
conditions which are rooted in technical systems
but cannot be understood when ignoring
the social function they serve
as socio-technical aspects.
We choose the term ``socio-technical'' to recognize
that social and technical aspects of the conditions
we describe here are interwoven.
The three socio-technical aspects
that we have categorized are \textit{complexity},
the \textit{software development process}, and \textit{safety \& responsibility}.

\subsubsection{\strongevidence{}~Complexity}

Most interviewees in our study
relate testing to complexity.
\reviewmark{They consider projects as complex,
when there are dependencies
between modular components
and the project is so big that
it becomes difficult
to have an overview of interacting components\groundedin{2}{0.125}.}
\code{DevelopingComplexPrograms} makes
testing more difficult.
Especially when testing
is not introduced early
in the project (\code{iterativelyReachingTestingMaturity}),
it can be challenging
to apply testing techniques (\code{strugglingWithComplexity}).
Complexity of projects
can then be perceived
as overwhelming (\code{OverwhelmedByComplexity})\groundedin{8}{0.5}.
Overwhelming complexity can however also motivate
testing efforts.
Testing is used to gain
overview\groundedin{10}{0.5}.
Interviewees also explain
that the social dimension of complexity
affects software testing.
For example,
the size of teams (\code{copingWithTeamSize})
has an effect
on how developers communicate
and how they are organized (\code{oversightInTeams}).
Interviewees tell us that
in large teams,
testing is used
to cope with the communication challenges of large teams (\code{reducingComplexityOfCommunication}\groundedin{6}{2}.

\revised{Whether complexity is perceived as a motivating
or limiting factor
depends on the concrete software project.}{2.7}
For example,
if sophisticated testing infrastructure,
resources and know-how are present,
complexity is less likely
to lead to overwhelming experiences.
Complexity then
positively influences
testing ambitions.
This interdependence
of complexity
and other factors
illustrates that
not a single factor
alone affects testing choices.
We propose Hypothesis~\ref{hyp:1}
on the basis of this analysis.

\subsubsection{\evidence{}~Software Development Process}

The way in which the process
of software development
is organised has an impact on how
developers approach testing.
For example, where iterative software development is embraced,
software testing can be considered a supporting technique
(\code{continuousFeedback})\groundedin{57}{1}.
The notion of \code{testingAidedAgility},
when testing works hand in hand with
iterative approaches.
\reviewmark{Testing also supports code reviews and
iterative design of software
(\code{TestingAidedCodeReview}\groundedin{2}{0.105}, \code{TestingAidedDesign}\groundedin{5}{2.5}).}
Testing can be used to establish a baseline (\code{AddingSolidGround})
from which developers can continue to iterate with confidence\groundedin{3}{1}.
However, testing can also be perceived as an inhibitor of agility.
Contributing test code might then be perceived as
adding technical debt and overhead\groundedin{1}{2.5}.
The assignment of roles
within the team also
conditions testing choices.
In agile teams
with \code{crossFunctionality}\groundedin{1}{1}
when developers are responsible (\code{DeveloperResponsabilities})
for quality assurance,
testing can take on an important role\groundedin{1}{1.5}.
It either motivates developers
to embrace the responsibility\groundedin{43}{0.5}
or it becomes a burden\groundedin{1}{2}.

\subsubsection{\strongevidence{}~Safety \& Responsibility}

Testing can act as \code{guardRails}\groundedin{3}{3}
which help them \code{preventingDamage}
and give developers a feeling of safety\groundedin{55}{0.75}.
Testing is done when that safety
and consequently confidence is needed\groundedin{8}{0.7}.
For example, when there is an awareness of
potential bugs being a threat to
oneself, other developers
or end-users (\code{TestingAidedConfidence}\groundedintwo{3}{1}{51}{4}).
An \code{awarenessOfResponsibility}
can be addressed with testing strategies.
Finally, providing a \code{testingSafetyNet} through testing
makes \code{failingInControlledManner} possible\groundedin{57}{2}.
In projects with high developer fluctuation
testing can be used to encourage
newcomers to contribute\groundedin{8}{0.2}.
\subsection{Affordances}
We take the term \textit{affordances}
from \citet{gibson_ecological_1986},
who defines that
affordances of the environment are
what it offers to humans or animals,
what it provides or furnishes,
either for good or ill.
The term implies the
complementarity of
the subject and the environment~\citep[$\S8\P2$ p.127]{gibson_ecological_1986}.
In our work, we categorise
perceivable circumstances in the environment of a developer
that offer something to them
as affordances.
For example, theoretically a tool determines an easy means of testing.
But whether and how developers
are able to perceive its value from their unique perspective
and use it depends on what the tool affords in its
unique relation with the developer.
Conditions categorised as affordances here
include how developers engage in
software development through available materials or constraints
(e.g., how does the application- and business domain afford testing to developers?).
The five conditions we have categorized under affordances are
the \textit{business \& application domain},
\textit{vision},
\textit{resource usage},
\textit{mandates}
and \textit{testing infrastructure}.

\begin{figure*}
	\centering
        \includegraphics[width=\textwidth]{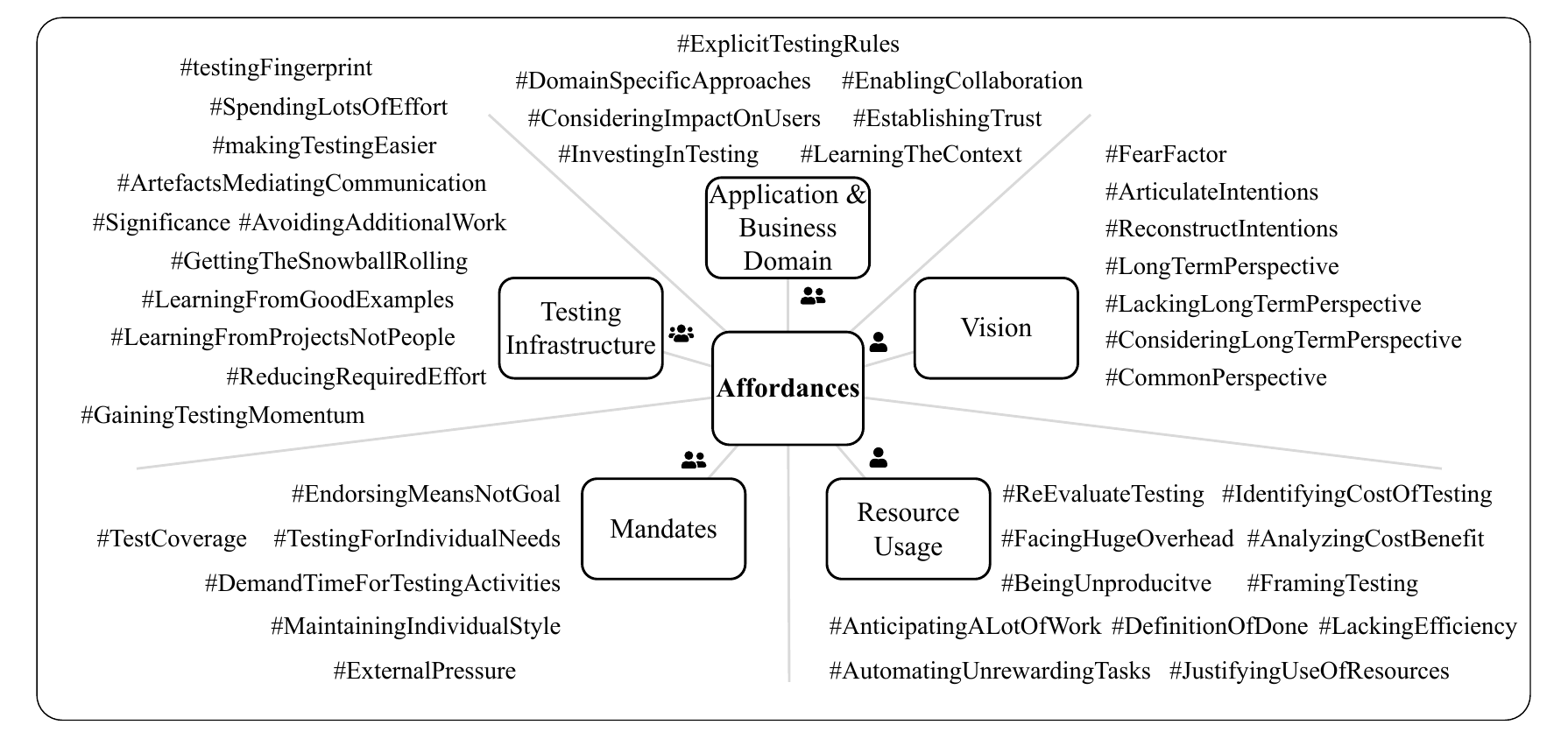}{}
        \caption{Diagram showing subcategories and
            the most prevalent codes which contributed to the forming of
            the analytical category \textit{affordances}.
            Through constant comparison of interviews through codes we identified
            conditions that influence when developers use testing practices.}\label{fig:affordances}
\end{figure*}

\subsubsection{\evidence{}~Business \& Application Domain}

The business
and application domain
in which a project is embedded
conditions how testing is done.
For example, for developers who develop
machine learning applications
certain aspects of testing
are less relevant than for developers working in other domains\groundedin{7}{0.5}.
Developers who had worked exclusively in those domains
may lack an awareness of techniques like unit testing.
Testing strategies are therefore subject to
\code{DomainSpecificApproaches}.
Software development projects are
also embedded in a business context
which determines
who the software is developed for (\code{ConsideringImpactOnUsers}),
the risks of introducing bugs
and juridical aspects\groundedin{10}{0.8}.
Governmental regulations
might necessitate a strict
quality assurance regimen for specific applications,
justifying how many resources
a project invests in testing\groundedin{6}{2}.
\code{InvestingInTesting}
can be motivated by other
business specific factors as well.
Consider a
free/libre open-source software (FLOSS) project
that welcomes contributions from anonymous developers.
Here testing can be an important means of \code{establishingTrust}.
Further, anonymous contributions influences how
testing expectations are
communicated (\code{ExplicitTestingRules}),
changing how visible it is to developers\groundedin{55}{2}.
Contribution guidelines
can make testing
expectations explicit, which
in turn has an influence
on developer's choices.
The absence of such explicitness
on the other hand
may require developers to pick up
implicit knowledge
through other means.
\code{LearningTheContext} of domain and business is necessary
in order to know when and how to test\groundedin{6}{3}.
\reviewmark{The usage of stable and freely available, common tools
(which is arguable more common in FLOSS-projects)
also contributes
to the long-term value of each contributed test\groundedin{2}{0.225}.}

\subsubsection{\weakevidence{}~Vision}

Considerations about
how a project will be used in the future,
how it is going to be changed and maintained
and who will be responsible for those tasks impacts
testing choices.
Long term planning
impacts the confidence
required by maintainers to
accept changes to a project\groundedin{55}{0.5}.
When developers expect that
they will inherit responsibility
for projects in the future,
testing may be perceived in a different light\groundedin{10}{0.25}.
When there is a high fluctuation of developers
the knowledge about what the purpose of source code is can get lost.
Testing practices can be a safeguard against
losing this important information
as tests can \code{articulateIntentions}\groundedin{43}{2}
and can therefore potentially be used to \code{reconstructIntentions}\groundedin{3}{1}.
The shared vision for a project
can motivate testing as developers are
\code{ConsideringLongTermPerspective} of projects
when evaluating whether to test\groundedin{49}{2}.
Testing contributions might even be considered
to have an effect on long term job security\groundedin{3}{4}.

\subsubsection{\weakevidence{}~Resource Usage}

\reviewmarktwo{Establishing and maintaining
a testing strategy
costs resources like working hours (time),
or technical investments
which developers are aware of (\code{identifyingCostOfTesting})\groundedin{4}{1.2}.}
Decisions about how resources
should be used
condition the choice of when to test (\code{analyzingCostBenefit})\groundedin{58}{1}.
\code{FramingTesting} goals
in terms of
its value can help developers
to prioritize testing (\code{justifyingUseOfResources})\groundedin{3}{2}.
Solely allocating more resources
does not necessarily lead to
more testing\groundedin{51}{2}.
Making adequate resources available
to developers needs to
go hand in hand with communicating the \code{significance} of testing
so that developers can change priorities
according to a common understanding of
the value of testing for a project (\code{ReEvaluateTesting})\groundedintwo{4}{1.002}{1}{3}.
A feeling of having the power
to contribute something significant
with testing is what affords
it in software projects.
Testing is avoided
when resources spend on
it are considered as \textit{wasted}
(\code{facingHugeOverhead},
\code{beingUnproductive},
\code{anticipatingAlotOfWork},
\code{lackingEfficiency})\groundedin{1}{2}.
Structuring work in such a way that testing
is not perceived as an \code{ExternalPressure},
but as a productive task,
for example by including testing in the \code{DefinitionOfDone}
can prevent this\groundedin{3}{2}.
Interviewees tell us they consider
testing a productive rather than a wasteful activity
when testing is framed positively.
Especially when testing can help developers with \code{automatingUnrewardingTasks},
it is appreciated even by those who usually perceive it as a burden\groundedin{1}{2.25}.

\subsubsection{\evidence{}~Mandates}

We categorize as testing mandates rules
or recommendations for testing which are
imposed on developers.
For example, rules that demand from a developer
that a certain percentage
of \code{TestCoverage}\groundedin{3}{0.5}
needs to be
maintained
with every code change.
Mandates
can
encourage
developers
to keep testing
in mind
when starting
a new project\groundedin{10}{0.3}.
Mandates enable developers to
plan and ask for investment of resources into
testing efforts (\code{demandTimeForTestingActivities}\groundedin{3}{2}.
However, mandates can also
effect testing choices negatively.
Taking autonomy away from developers,
requiring them to use testing practices
can impose a burden\groundedin{43}{1}.
When testing is not perceived as worthwhile by developers
they might work around the mandate,
developing useless tests\groundedin{1}{1.5}.
Keeping testing efforts to a minimum
in that way can negatively influence how
testing is perceived by collaborators
as they interact with and learn from the \code{testingSignatures}
of their collaborators.
Mandates introduce
an \code{externalPressure}
which can even punish those who follow
the mandate in a meaningful way\groundedin{1}{1}:
mandates and guidelines change the relation
between developers and testing practices in projects\groundedin{1}{1.5}.

\subsubsection{\strongevidence{}~Testing Infrastructure}
\reviewmarktwo{Visibility and usability
of testing infrastructure
like test cases enable
developers to copy and extend,
making the process
of developing tests
easier (\code{makingTestingEasier})\groundedin{4}{1.005}.}
Artefacts
also act
as symbols,
signifying the importance
of testing (\code{ArtifactsMediatingCommunication})\groundedin{43}{2}.
They communicate
the \code{significance} of testing
in a project
and exemplify the measure of testing
which is used by collaborators.
We argue that developers
leave a \code{testingFingerprint}
in projects
from which collaborators learn (\code{learningFromGoodExamples})\groundedintwo{45}{1}{53}{1}.
We were told by our interviewees that
this indirect communication between
developers to pass on testing knowledge can
be more impactful than interpersonal guidance:
\reviewmark{Being exposed to testing artifacts and seeing the \code{testingFingerprint} of someone
can start a process of reflection that can lead to a change in how
testing is perceived (\code{learningFromProjectsNotPeople})\groundedin{2}{0.15}.}

The lack of testing infrastructure
inhibits testing efforts (\code{AvoidingAdditionalWork}, \code{findingEfficientSolution}).
Especially in complex projects,
taking the first steps to establishing
a working testing infrastructure can
be a daunting task\groundedin{8}{0.5}.
\reviewmark{\reviewmarktwo{Building up an initial suite of tests
or \code{refactoringForTestability}\groundedin{4}{1.5}
can take a lot of effort when the project was build with
no testing in mind (\code{spendingLotsOfEffort}\groundedin{2}{1}).}
To establish software testing in a project
a challenge for developers often is
do exactly that
to establish a momentum for testing efforts
(\code{GettingTheSnowballRolling},
\code{gainingTestingMomentum})\groundedin{2}{1}.}
\subsection{Dogmatic Perspectives}
\begin{figure*}
	\centering
        \includegraphics[width=\textwidth]{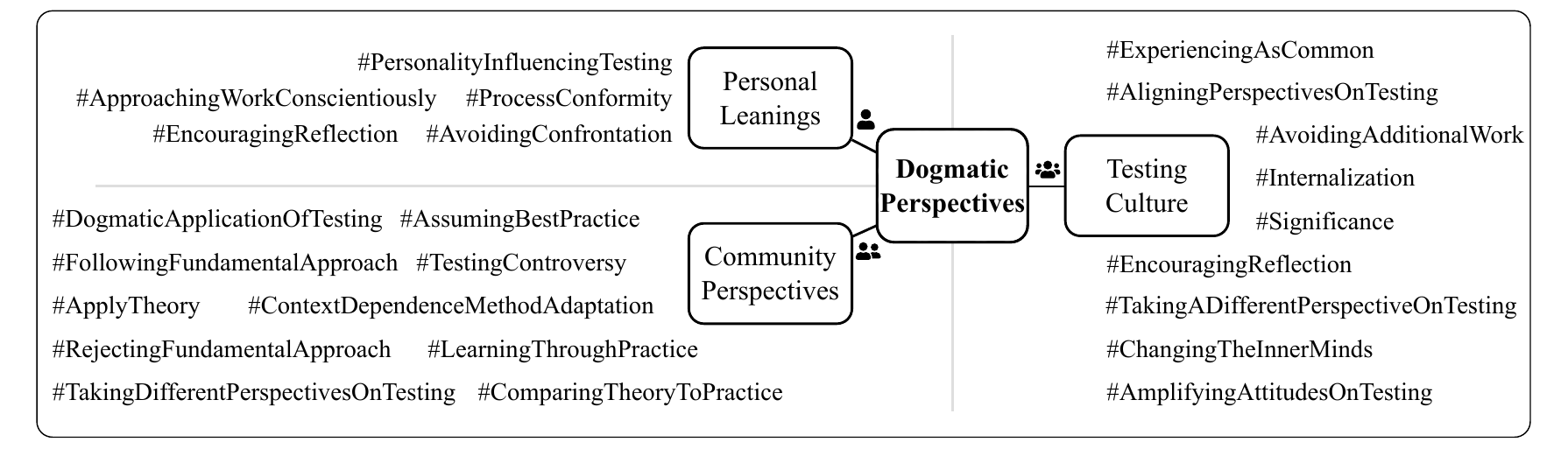}{}
        \caption{Diagram showing subcategories and
            the most prevalent codes which contributed to the forming of
            the analytical category \textit{Dogmatic Perspectives}.
            Through constant comparison of interviews through codes we identified
            conditions that can create dogmatic perspectives which have an effect on how testing
          is used in projects.}\label{fig:dogma}
\end{figure*}
\reviewmark{\reviewmarktwo{Socio-technical aspects
and affordances alone cannot explain why
some developers and even teams
aspire to testing with conviction
even though circumstances
are completely disadvantageous\groundedintwo{2}{2}{4}{1.005}.}
We find that adoption
and adaption of testing practices
goes hand in hand with
a culture change within projects\groundedin{2}{1}.}
Strong opinions of developers
which can be perceived as facts by collaborators
contribute to this change in culture\groundedin{1}{3}.
Such dogmatic perspectives
can motivate or discourage
developers and their teams to test. 
in testing.

Conditions for testing which
we categorised as dogmatic perceptions
prime how a developer perceives testing.
The three conditions we have categorized under dogmatic perspectives are
\textit{testing culture},
\textit{community perspectives}
and \textit{personal leanings}.

\subsubsection{\strongevidence{}~Testing culture}

The interviews we conduct provide strong evidence
that software projects develop a testing culture
that influences how each individual developer
relates to testing
within the scope of
that project.
Testing culture is ephemeral in the sense that it
is often beyond the grasp of formal documentation,
resists conscious planing and is therefore rarely written down.
Developers do however acknowledge testing culture
and even explicitly attribute some of their
choices to it\groundedin{51}{1}.
Testing culture reflects
the collective understanding
of what to test,
how much to test
and how to test it.
\reviewmark{The culture within a team
establishes the
boundaries of testing practices (\code{alingingPerspectivesOnTesting})\groundedin{2}{2}.}
\reviewmarktwo{It establishes what
is considered
to be \textit{common} (\code{ExperiencingAsCommon})\groundedin{6}{2.5},
and what is considered unnecessary, wrong, or
even crazy\groundedin{4}{1.2}.}
\reviewmark{By \code{amplifyingAttitudesOnTesting},
the testing culture in a team
can even declare a project
\textit{untestable},
freeing developers from the burden
of testing\groundedin{2}{2}.}
Change in testing culture
is not created by mandates or simply
by changing other affordances or socio-technical aspects.
It requires what
interviewees describe
as a process of \code{changingTheInnerMinds}\groundedintwo{5}{1}{10}{0.85}
of collaborators.
\reviewmarktwo{We find that this process
is stimulated by \code{encouragingReflection}
and by \code{learningThroughCollaboration}
which enables \code{takingDifferentPerspectivesOnTesting}\groundedintwo{4}{1.2}{47}{1}.}
Changing the constellation of a team
by for example adding a new developer
is therefore likely to alter its testing culture\groundedin{5}{4}.

\subsubsection{\evidence{}~Community Perspectives}
Beyond the scope of software projects,
developers learn
from a potentially global community
who advocate or condemn specific testing practices.
Those community perspectives
are represented on Q\&A platforms like \stackoverflow{}~\cite{swillus_sentiment_2023},
interactive forums, or at practitioner conferences.
Community perspectives
and opinions also reach developers through gray literature
like books, magazines, or blogs~\cite{garousi_when_2016}.
\revised{Perspectives on testing shared through any
of the aforementioned ways
are detached from the concrete projects
that a practitioner is participating in.}{2.7}
A university program
or a blog post
cannot account for
the specific conditions
in real-world projects.
Reconstruction of knowledge is required
in order to apply it to projects\groundedintwo{4}{1.002}{5}{2}
\reviewmark{Teaching and learning testing
is challenging because
translating theory to practice
in the context of its application
can be difficult \code{ContextDependentMethodAdaptation}\groundedin{2}{0.13}.}
\revised{When translating
from theory to practice (\code{comparingTheoryToPractice}, \code{learningThroughPractice})
in a complicated situation
does not happen,
\code{ApplyingTheory}\groundedin{57}{1.5} can lead to inflexibility
and \textit{over-fitting} of methods.}{2.7}
\revised{\reviewmarktwo{Learning \textit{theoretical}
community perspectives
can lead to
dogmatic perspectives
that impact testing choices,
especially when there is no process of translating
theory to applied practice (\code{followingFundamentalTestingApproach})\groundedin{4}{1.5}.}}{2.7}
It can
also create resistance
against testing practices
to be used for example when it
opposes their application\groundedin{43}{1.5}.
Community perspectives add normative value
to testing practices.
\reviewmarktwo{We see examples of this tendency
when developers
regard specific practices as ``normal''
best-practices (\code{assumingBestPractice})\groundedinthree{4}{1.8}{8}{0.15}{53}{1}.}
\reviewmark{On the other hand,
when interviewees elaborated
on nuanced perspectives (\code{takingDifferentPerspectivesOnTesting}, \code{rejectingFundamentalTestingApproach})
we were warned explicitly
that the opinion could be controversial (\code{TestingControversy})\groundedin{2}{0.11}.}
Community perspectives,
when described and discussed
in a way that is removed from
their actual application,
can create dogmatic attitudes
(\code{DogmaticApplicationOfTesting}).
\reviewmarktwo{Dogma which is followed by individuals
can then stand in the way of reasonable testing approaches\groundedin{4}{1.5}}

\subsubsection{\weakevidence{}~Personal \textit{leanings}}

Reflecting on experiences in different companies
and recalling the polarizing opinion of a colleague
who thought that testing was useless,
one interviewee argues explicitly that decision-making
regarding testing is dependent
on factors that go beyond software development
and are related to individual character (\code{PersonalityInfluencingTesting})\groundedin{1}{3}.
Apart from that we are not able to pin down how character traits
map to attitudes on testing through the analysis of interviews,
but the notion
that there might be a correlation
is mentioned by multiple of our
interviewees.
\reviewmarktwo{For example they hypothesize that
it is easier to establish testing practices in projects when
developers are conscientious\groundedin{4}{1.7}.}
From the evidence interviewees
shared with us in interviews
and from the way in which they
describe themselves and their own approach to
development and testing we find
the remark that the perceived importance of testing
maps to personal leanings plausible.
For example, specific
personal leanings or character traits
might causate with \code{processConformity}
or a tendency to shirk away from mandatory testing
tasks.
\section{Discussion}\label{sec:related_work}
We conducted interviews with developers
to find out how they relate to software testing practices.
Developers shared stories with us which
we analyzed systematically to develop
a novel theoretical perspective on testing strategy emergence.
 We argue, that testing ideas and solutions are
 constructed, imagined, discussed and reflected through
 social interaction.
 Developers reflect about testing in formal and informal
 settings to understand why it should (not) be done in a certain way.
 Through their testing contributions
 they translate those reflections
 into more permanent artefacts
 like source code or testing infrastructure.
 We theorise
 that the imagination,
 implementation, and improvement of
 testing strategies is driven by
 the effect of
 conversation (\code{TestingEchoes}) on
 collaborative
 development of testing artefacts (\code{TestingSignatures})
 and vice versa.
 Additionally,
 socio-technical and organizational factors
 which empower developers to use testing techniques (\code{testingAfficacy})
 are entangled in this mutual relation.
 Testing contributions alter the conditions
 of their development:
 For example, the creation of an initial set of unit-tests
 makes the development of tests easier as they can be copied.
 We argue that testing strategies emerge from a recursive process.

In this section we
first situate our work
in the body of knowledge
of our field and present novel vantage points,
relating our theoretical contributions
to literature beyond the scope of software engineering research.
We then discuss why our findings are important to both software engineers and researchers
in software engineering field.
Finally, we motivate future work
and critically reflect our own work to extend
and challenge the views we present.
\subsection{Related Work}\label{sec:form}

Whether developers are willing to adapt
their behavior
to follow a software development methodology
is influenced by the perceived usefulness of a method,
the social pressure in their teams,
the compatibility with their current work and responsibility,
and organizational mandate~\citep{hardgrave_investigating_2003}.
We find that social- and human factors
also play a big role in the case of software testing.
Software testing is not a merely technical phenomenon, but
a social experience that can even reveal a wide range of deep
emotional connection between developers and tools~\citep{evans_scared_2021}.
Developers use testing when they see it fit
and their perspective is shaped
by material realities
(e.g., source code under test and available infrastructure)
and based on how testing is imagined,
discussed and valued
between individuals of a project (e.g., testing culture).
Our findings align with findings of \citet{hardgrave_investigating_2003},
who argue that
social pressure has a
bigger effect on choices than organizational mandates.
Even when developers think that a methodology is useful,
social pressure can make them resist change.

Further, our findings suggest that the likelihood
of adoption of testing methodologies increases,
when
developers
are allowed to adapt them.
As Conway already proposed in 1967,
introducing variation requires
granting the members of an
organization autonomy~\citep{conway_how_1968}.
The importance of developer autonomy is also
mentioned in gray literature, for example
in the agile manifesto.
Within their projects,
developers should be enabled
to find a way that fits their teams needs.
Both Conway and the agile manifesto
argue that developers need to be trusted to get the job done~\citep{kent_beck_manifesto_2001}.
This is also suggested by \citet{rooksby_testing_2009}:
Being able to deal with various arising contingencies
is what makes a plan to test work;
developers should be allowed to
deviate from a plan in order to sustain its spirit.
Adding to what \citet{rooksby_testing_2009} and \citet{hardgrave_investigating_2003}
state, our findings suggest that being able to
adapt methodologies,
testing culture
and technical material (e.g., code and infrastructure)
is what makes a plan to test work.
Our interviews provide us with strong evidence
that the function of testing
(what it tries to achieve and how it does that)
needs to be aligned or follow the form of
teams (who they are and how they collaborate with each other)
to be effective.
One interviewee directly pointed us
to Melvin Conway's law~\citep{conway_how_1968}, which
establishes this link between
form and function\groundedin{47}{2}.
Conway suggests that:
\textit{Organizations which design systems
    (in the broad sense used here)
    are constrained to produce designs
    which are copies
    of the communication structures
    of these organizations.}
Given a team's organization, Conway argues,
there is only a class of design alternatives
that can be effectively pursued by the team.
In the case of testing
we find additional factors
which seem to steer and constrain the design of systems.
Testing strategies are developed using available material (infrastructure and code).
Their design is influenced by a team's constellation (including testing culture),
and the business context for which it is developed.
We name those and other concrete factors in \Cref{sec:conditions}.
The ever-changing organizational
landscape of projects,
an ever-changing assortment
of technologies and an ever-changing diversity of people
leads to a plethora of development teams
and therefore to a big variety of testing strategies.
Consider the case of open source projects,
where the means of communication and the means of production
of artefacts are radically different
from non-open source software projects.
When open source projects welcome one-time contributions of anonymous developers,
mechanisms to establish trust between maintainer and contributor
and the limitations of their remote communication will be reflected
in the testing strategy (e.g., how explicit it is).
The means of production and
its social- and organizational foundations (open-source),
including subjective perspectives of contributors,
determine the function of testing.

In \Cref{section:theory:recursive} we argue,
that testing strategies
are not deliberately placed or designed.
Their design is stochastic
because it is
influenced by a complex set of interdependent conditions
(e.g., testing infrastructure, testing culture).
\revised{%
Conditions we identify are similar to those
proposed by the technology
acceptance model in its most current version (TAM3)~\citep{venkatesh_technology_2008}
(e.g., Application and Business Domain $\rightarrow$ Job Relevance,
Subjective Norm $\rightarrow$ Testing Culture).
TAM,
the most widely adopted acceptance behavior theory
in the field of software engineering research~\citep{borstler_acceptance_2024, lorey_social_2022},
highlights that conditions impact the perceived usefulness
and perceived ease of use of technologies.
We similarly construct the concept testingEfficacy.
Additionally, in \Cref{sec:theory},
we describe how
knowledge and opinions about testing
spread through multiple channels (testingEchoes, testingSignatures).
The theory of diffusion of innovation (DOI)~\citep{rogers_diffusion_2014}
explains the role of such \textit{diffusion arenas} in impacting technology adoption.
However,
the results of the present work
raise doubts about the applicability of both theories
in the context of software testing.
We argue that an application of TAM
risks taking a view that is too narrow~\cite[see][]{lee_technology_2003}.
We argue,
that testing strategies are strongly impacted
by collaborative dynamics which are underrepresented in TAM~\citep{salahshour_rad_information_2018}.
Further, our theory explains
that conditions for testing
and testing strategies
continuously change
through a recursive process.
The historic development of a project
has an impact on testing adoption and testing strategy adaption.
However, contrary to our theory,
DOI postulates that
historic developments do not have an effect on the progression of innovation adoption.
This incompatibility with DOI
has also been identified by \citet{lyytinen_whats_2001}
for other complex information systems.}{2.1}

As history is never fully knowable,
the causes of testing decisions and their effects
on the testing strategy
are also never fully knowable.
Instead of being deliberately placed or designed,
testing strategies
therefore emerge from an interplay
of conditions
which are being introduced by different
actors to reach goals that are not always clear.
Testing strategies, we argue continuously develop and evolve
collaboratively from \textit{within}
projects in a way that might not be sensible to an outsider.
Whether the function of testing serves its
intended goals
and what stands in the way of improvement cannot be judged without
a team's sensibility.
\revised{%
Similarly, \citet{rogers_diffusion_2014}
reflects on the export-driven, top-down approach
that theories like DOI take.
As research mostly focuses on identifying and removing
external factors that stand in the
way of adaption (e.g., TAM, DOI),
inside-out, ground-up approaches are often overlooked.
Positive deviance~\citep{marsh_power_2004},
which is recommended by \citet{rogers_diffusion_2014} as an alternative,
acknowledges and embraces local knowledge
to find solutions
that work for communities and fit their culture.
Our findings suggest that such inside-out
approaches are especially suited to support
projects in their testing efforts.}{2.6}

Developers learn from interacting with code of others (signatures)
and discuss and reflect what they experience with each other (echoes).
Reflective learning
seems to be a driving force
in the process of
testing strategy adaption and has been studied extensively
by \citet{schon_reflective_1983}.
He argues that practitioners
are often confronted
with situations in which
there is yet
no \textit{problem} to be solved
because the goal
in a problematic situation
is not always clear.
\textit{Problems} he finds, are constructed
by practitioners from the material of problematic situations
which are puzzling, troubling and uncertain.
Practitioners clarify problematic situations
through non-technical processes so that
the ends to be achieved can be framed and organized.
Only then the possible (technical) means
to \textit{solve the problem} are identifiable~\citep[$\S\RN{1}.2.3\P\text{10-11}$ pp.40-41]{schon_reflective_1983}.
What we describe
as a process that alternates
between the generation of transient,
non-documented and spontaneous \code{testingEchoes}
and the creation of artefacts and thereby \code{testingSignatures},
is similar to what
Sch{\"o}n calls \textit{Reflection-in-Action}.
Reflection focuses interactively on
the outcomes of action,
the action itself and the
intuitive knowing that was
implicit in the action~\citep[$\S\RN{1}.2.4\P\text{23}$ p.56]{schon_reflective_1983}.
Faced with a yet puzzling situation
a developer who is not yet familiar
with the ins and outs of testing
begins with experiments.
They re-use
existing test cases (\code{testingSignatures})
by copying, pasting and modifying~\citep{aniche_how_2022}.
Succeeding and yielding good results
they are reflecting-in-action,
considering what it was that lead to the result.
Reflecting their intuitive knowing and the norms on which
it is based can then be an
impulse to discuss with collaborators (\code{testingEchoes}).
Interviewees refer to this process when
they say that it was not people but the projects
to which they contributed that taught them testing\groundedin{2}{0.15}.
Sch{\"o}n's theory can also account for
what interviewees describe as an \textit{overfitting}
of ideas~\citep[$\S\RN{1}.2.4\P\text{40}$ p.60]{schon_reflective_1983}.
Developers with \code{dogmatic-perceptions}
who have \textit{over-learned} a specific technique or method
become unattentive to phenomena that do not fit the categories
of their knowledge.
They lose their ability to adapt, something which is considered important
for testing~\citep{hardman_testing_2025}.
Sch{\"o}n argues that reflection,
especially when related
to problematic situations
of the past,
is necessary to overcome this over-fitting.
Practitioners do not only reflect and learn in-action,
but also through post-mortem analysis and discussions.
Conscious and unconscious
reflection of past situations alters
how developers interact with code,
how norms and appreciations are involved in that interaction,
how the interaction is situated in the larger institutional context
and finally how this interaction constitutes a collaborative
effort~\citep[see][$\S\RN{1}.2.4\P\text{42-45}$ pp.61-62]{schon_reflective_1983}.

Consider, stories about \textit{armageddon-bugs},
which, as one of our interviewees reports\groundedin{58}{1}
make their way into
the on-boarding process
of new developers
and stimulate
reflective discussions
about the benefits and pitfalls
of testing methods.
Post-mortem anecdotes create \code{testingEchoes},
which motivate testing strategies
and generate
\code{testingSignatures}.
In the form of source code or
infrastructure
\code{testingSignatures} become representations
of \textit{how} testing can be done.
The process of collaborative reflection
and reflection-in-action as
developers engage
with testing artefacts
here is not a guarantee to prevent the next
disaster.
Neither is it a goal in itself.
Instead, analog to Sch{\"o}n's argument,
we theorize that the adaption of testing strategies
is a collaborative means to
act on an uncertainty that arises
from the particularities of practice
in a unique and ever-changing environment (RQ2, RQ3).
Engaging in testing practices facilitates
and is sometimes even guided by
reflections about misconceptions
and mistakes that arise in volatile environments.
We argue that opinions about testing are
connected to those reflections which can often be attributed
to concrete, sometimes emotional events (RQ1).

\subsection{Implications}\label{sec:implications}
Software testing is seen
as a means to increase quality and efficiency,
and as a tool that can
steer the technical design of software projects~\citep{santos_family_2021}.
The results of this work
suggest that
software testing practices
have an impact that goes beyond technicalities.
Like other scholars~\citep{rooksby_testing_2009, garousi_when_2016, evans_scared_2021},
we argue that
one can only understand
the impact and potential of testing practices,
by giving attention
to the socio-technical
and material reality
in which it is applied.
The theory we propose
which is grounded in the data we gathered
and aligned with the work of
\citet{schon_reflective_1983}, \citet{conway_how_1968},
and scholars in the field of software engineering~\citep{rooksby_testing_2009, wiklund_impediments_2017, evans_scared_2021},
shows that software testing practices are strongly
influenced by an entanglement of social and technical factors
which is unique for every project.
Technical factors like the presence of testing infrastructures
and socio-technical factors like the development process followed by developers
inform and condition the testing choices developers can make.
We argue that whether any testing practice is useful depends
on the context in which it is practiced
and how well it is adapted to this context.
Effective testing strategies
are not only enabled by
measurable technical- and organizational circumstances.
They are envisioned through and conditioned by
human and social needs of developers.
As the goal of testing is often unclear
and as the practice of testing resembles
not a linear but a stochastic process,
the capacity of developers to
collaboratively reflect on their
needs, their knowledge and their actions
can be crucial for its success.
We argue that any deliberations
by researchers, developers and managers
to improve testing should consider the
non-technical factors we name in our work.

Our empirical research which investigates real world experiences
challenges us to produce outcomes which are relevant and
actionable.
Claiming that we understand what is going on,
we should be able to identify issues and
suggest solutions.
However, investigating real world experience
also comes with a realization that
there are no simple solutions to suggest.
Through our conversations with developers,
we learn
that the best solutions
are most often suggested by developers themselves
as they are best suited to identify the problems that need solving.
They are better suited than we are
to find out
what they need to do in order to reach their goals.
We therefore want to suggest methods
that can help developers in the \textit{process of finding out}
what their issues and solutions are.

\subsubsection{Socio-technical Construction of Testing}

We encourage developers to consider
that software testing
is not a solely technical matter.
Its value and applicability
is socially constructed.
This means that a testing strategy
is not just a composition of
testing methods and technical implementation.
Instead, its value is constructed
through a collaborative process
by developers and other stakeholders.
Consider testing mandates that require testing for every code commit.
Mandates justify a developers' investment of time into testing
and can build confidence among contributors.
High code coverage suggests
that changes are less likely to break
existing functionality unknowingly.
Demanding high code coverage
can thereby create an
environment where newcomers feel welcome
to introduce changes,
but it can also lead
to developers adding useless tests just
to reach mandated metrics,
making the source code more difficult to maintain
and less intuitive for newcomers.

\revised{Not only the value of testing,
but the knowledge about it is
socially constructed.
Testing knowledge goes beyond technicalities
and becomes valuable through interaction and sharing.
Embracing a collaborative attitude
and fosters what
is needed to move a project's testing efforts forward.
Considering all voices
and allowing capabilities and technological
strategies to evolve together prevents
the testing strategy from disintegrating or becoming an end in itself.
For example, changes in a team's constellation
when an individual joins or leaves
can be seen as a valuable opportunity to re-align
testing strategies and technologies.
We encourage developers'
prioritization of learning and reflection
over the pursuit of flawless technical solutions.}{1.17}
This requires developers to be able to question
established views.
Wether from research (including our own),
online communities, or blogs.
We suggest that developers' focus should
not only be outward but placed on
knowledge that already exists in the organization.
We suggest considering approaches like Positive Deviance
which promote inside-out processes~\citep{marsh_power_2004}.
Providing developers opportunities
to reflect about their development experience
can surface issues or needs related to testing.
Code-reviews,
formal meetings like retrospectives,
and informal conversations can be opportunities
to encourage each other to reflect
on what is done in terms of testing, and why it is done.
The goal does not need to be finding a perfect solution
but simply to exchanging stories and experience.

\subsubsection{Materiality of Testing}

Artefacts that
are related to testing practices, i.e.,
the materiality of testing,
shape how testing
is used, perceived, valued
and approached.
This includes artefacts that are created through testing (e.g., test code)
artefacts which support testing practices (e.g., CI/CD infrastructures)
and not only the artefacts which are subject to testing (e.g., source code that is being tested).
We argue that the absence, presence and availability of artefacts
causes effects
that go beyond
the functional goals of a project.

First, the availability of infrastructure
in the form of tools and existing tests
can make testing easier and more approachable.
For example, working test code
can be used as a template,
being copied, modified and evolved.
Improving and maintaining testing infrastructure
can \textit{get the snowball ball rolling}
as one of our interviewees puts it.
Creating momentum through
infrastructure improvements can cause
an \textit{avalanche} of testing contributions
as each contribution makes testing easier
which motivates
more contributors to join in. 

Second, software testing artefacts
with which developers interact
communicate the significance
and preferred manner of testing
(the testing culture of a project).
They teach developers
how to test in a project.
Testing efforts
can actively make use
of this potential.
One way to do this is to develop code
that explicates intentions.
Concretely,
when naming test cases,
describing code doc-strings or comments
or when engaging in code reviews.

Third,
testing infrastructures affect
the (social-) interaction of developers.
For example, the results of a testing suite
primes the outcome of code reviews~\citep{spadini_test_driven_2019}
 which can potentially make the process more objective.
As this is not always desirable,
planning of infrastructures should consider
what the needs of developers in a project are.

\subsubsection{Implications for Research}

Researchers in the field of software engineering
have pointed out that
software testing is a cooperative process.
It has organizational~\citep{martin_good_2007},
psychological~\citep{garousi_survey_2013},
and social facets~\citep{rooksby_testing_2009}.
Others emphasize that
more insight into
the human experience of software engineering~\citep{sharp_role_2016}
and software testing in particular~\citep{evans_scared_2021} is needed.
Situating our work in the body of
knowledge available to us,
we identify a research gap
that lies beyond
technology-focused investigations
of socio-technical aspects of software engineering.
The gap concerns the effect
of material on the social worlds
in which software is developed (new materialism)~\citep{fox_new_2020}
and works that concern the social construction of technology (SCOT)~\citep{pinch_social_1996}.
Concretely this research gap
concerns two broad questions:
\begin{enumerate}
    \item How do material realities (software artefacts and development infrastructure)
        affect and facilitate social needs, interaction and the culture of projects?
    \item How does the transient, social experience of developers and their imaginations,
        which are not visible in artifacts, translate to choices?
\end{enumerate}

The gap exists because
contributions in the field of software engineering
and particularly software testing,
even when aiming to investigate
\textit{socio-technical} aspects
are often geared towards technical phenomena
through their choice of method or dataset.
For example,
\citet{wiklund_impediments_2014}
investigate forum posts
and conclude
that the impediments in testing are
of a technical nature.
We argue,
also on the basis
of our previous work~\citep{swillus_sentiment_2023},
that a forum or Q$\&$A platforms
is a very limited medium
to investigate non-technical issues.
\revised{
  \citet{evans_scared_2021} report that
  developers voice non-technical,
  even emotional issues with testing practice
when asked about it in-person~\citep{evans_scared_2021}.
Later, Wiklund also argues that
automated testing impediments
are for a large part of an organizational nature and
concern the human and social aspects of the lived experience of developers
as much as the technical details of it~\citep{wiklund_impediments_2017}.
As a result of those findings, \citet{evans_scared_2021}
develop the concept of Testing Experience (TX)
to emphasize that investigations into how to
improve practice need to go beyond what makes
testing efforts technically successful.}{1.7}
Works like those of \citet{wiklund_impediments_2017} and \citet{evans_scared_2021}
underline the
importance of socio-technical factors,
but concrete factors are rarely named
and their effect on testing is not explicated.
For example, we know that
training highly skilled developers
who excel in testing
requires teaching them soft skills~\citep{sanchez_gordon_beyond_2020},
but we do not know yet
why exactly these soft skills are needed.
Similarly,~\citet{garousi_when_2016}
propose a decision-making guideline for the
adoption of testing methods in software projects.
Their work provides a comprehensive overview
of factors which influence
software testing practices.
They conclude that both human and organizational
factors need to be considered when making decisions.
But the factors they name
(e.g., skill level, lack of support, resistance)
only slightly concern the lived experience of developers
focusing in greater details on technical aspects.

Testing, it appears,
is mostly regarded as a technical practice that
leverages technical advantages which are at most
influenced, sometimes conditioned
by social and human factors.
\revised{
An evaluation of testing literature (including gray-literature)
to develop testing guidelines~\citep{garousi_when_2016}
did not identifying any link between the social needs of
developers and testing.
Only recently,
the concept of Testing Experience (TX)
has been established~\citet{evans_scared_2021}
to describe testing as a practice that
goes beyond technicalities.
Like Evans et al. we also find
that testing is a multi-dimensional practice.}{1.7}
Reducing its many dimension to a quantifiable cost and benefit analysis
risks ignoring many of the benefits it can have for developers.
We argue that a purely technical framing
can discourage decision-makers
to engage with developers
in a meaningful way to find out
what the true, non-quantifiable value of testing is for them.

We hypothesize that
testing constitutes a complex and dynamic system of non-linear processes,
that is not only reflecting a projects'
culture but also facilitates the human needs of developers.
By not researching these human- and social factors
we miss out on the depth that the topic of software testing offers.
For example, to the best of our knowledge
the effect of testing practices
on social and human conditions
has not yet been researched.
Research only seems to consider the opposite perspective,
that testing is influenced by human and social factors.
Researching and proposing
new technological solutions
or praising particular methods
without addressing their implications
on the social- and human experiences
risks leading
to what we identify in our work as dogmatic views.

\revised{Works in the field of software engineering often highlight
that social aspects of testing are indeed important
but they do net tell us why.
They leave us hungry for more.
Taking the technical perspective to illuminate
whether something
has a social element
is not sufficient.
We would like to see researchers take
a socio-technical perspective on practice
from the start
to illuminate how the social- and human elements
affect developers.
Aligning with suggestions of \citet{whitworth_social_2009},
we argue that taking this perspective
will illuminate which social and human
needs are facilitated by testing approaches
and how these needs translate to technical requirements.}{1.17}
To make a concrete example:
we know that in order to be a good tester
one needs to be a team player~\citep{sanchez_gordon_beyond_2020}.
We should ask what this finding tells us
about software testing as a socio-technical practice.
If you need to be a team player
to be good at testing,
why exactly is testing requiring team work?
What exactly does this team work ask of developers?
In the following section we suggest
research questions that emerge from our own work
and address the research
gap we identify.
\subsection{Further work}\label{sec:future}
As discussed in the previous section,
further work could explore the following research questions:
\begin{itemize}
    \item[\textbf{RQ}] Which social and human needs can testing practices facilitate?
    \item[\textbf{RQ}] How does (an absence of) testing practice effect human- and social needs?
\end{itemize}
Not only can answering those questions
teach us more about the motivation
for developers to employ testing.
With our exploratory approach we
identified analytical categories
and constructed a theory that explains
why developers employ testing strategies.
Further investigations into the categories
for which we found strong evidence
are likely to reveal insights that
can be translated to
guidelines for developers.

Concerning the theory that we propose,
we recognize the potential for extensions and refinement.
Concretely, we propose mixed-method investigations
of the connection of what
we call \code{testingEchoes} and \code{testingSignatures}.
\begin{itemize}
  \item[\textbf{RQ}] How are transient impulses to use testing translated into artefacts?
  \item[\textbf{RQ}] How are testing artefacts generating transient impulses in projects?
\end{itemize}
Research could compare the material reality of testing to what developers say about it in interviews
to understand how their relation to testing is impacted by both artefacts and transient impulses.
Further,
the effect of artefacts on developers
could be researched through experiments, interviews or observation.
Recent studies have already shown that think-aloud experiments are
successful in revealing how developers consider testing artefacts on a technical level~\citep{aniche_how_2022}.
\section{Critical reflections and threats to validity}\label{sec:reflections}
Our systematic analysis of 19
interviews with software developers
provided us with new insights
and an interpretive theory into how practitioners
relate to software testing.
Within the STGT framework which we use for our investigation,
we take a constructivist stance.
We do not aim to find an objective truth;
instead, we aim to describe what is common to and true for
various observers.
The theory we propose is therefore
not an objective representation,
but aims to explain and predict
phenomena which influence developers'
subjective experiences.
We now critically discuss
the validity of our findings
with respect to
this constructivist stance we take.

\subsection{Credibility -- Internal validity}
As our study relies exclusively on interviews,
we identify multiple threats to internal validity
related to completeness and confidence in
representing the participants' experiences.

First, we identify the risk of biased data collection.
Most of our interviewees
were recruited through convenience sampling,
\revised{which introduces
a risk for biases as
the relation of interviewer and interviewee
might influence the conversation.
To mitigate,
we avoided recruiting individuals which
were known to the researchers through earlier collaboration.
4 out of 15 interviewees
match this criteria, one of which was also associated
with the institution of the researchers when the interview was conducted.
Further, also recruited interviewees
through other means.
We reached out to a broad international
audience by inviting \stackoverflow{} users
to interviews.
We also recruited developers
in everyday situations (e.g., developer conferences
or train rides).}{1.4}
In addition, participants share their experience
on the basis of their role or perceptions of what
the researcher wants to hear.
To reduce the effect of this positionality on
the results of our work,
we recruited developers
from multiple countries
who have different roles
in various companies
situated in several industry sectors.
Related to this,
we identify the risk that interviewees
provide socially desirable answers
or that they are reproducing dominant or common discourses
instead of offering insights which are rooted
in their own lived experiences.
To mitigate this risk
we used Charmaz' techniques
for semi-structured interviews which
encourage interviewees to reflect and
re-contextualise ~\citep[$\S4\P1$ p.85]{charmaz_constructing_2014}.
Taking a neutral stance, only nudging interviewees
to go deeper in their reflections,
we reduce the likelihood of influencing participants.
We follow a systematic
guideline for the construction
of (semi-structured) interview guides
by \citet{kallio_systematic_2016}.
For transparency, we make the interview guide available~\citep{swillus_interview_2026}.

Second, errors in translating audio recordings to transcripts
pose a threat to validity.
Transcription risks losing
important nuances like emphasis in voices,
mimic, etc.
We mitigated this threat by
manually transcribing interviews that
were conducted during the first stage of
data collection and analysis (the first 10 interviews).
For all other interviewees, we
automated transcription, but
manually reviewed generated transcripts.
Nuances and non-audible hints
that were noted by the interviewer
during interviews
were added to transcripts
during manual transcription
or transcript reviewing respectively.

Third, we identify the risk of
misinterpreting participants' views.
We mitigated this risk by
engaging in a number of systematic reflective practices.
Throughout the whole process of data
gathering and analysis, we wrote and compared
analytical memos which reflect
thoughts, assumptions, and potential biases.
Complementary to memo writing
we used practices like diagramming
and clustering to explore different perspectives
on the collected data.
Prolonged engagement with the data
through iterative data collection and analysis
increases our confidence in
our analysis; through continuous refinement we
ensure that all theoretical explanations are firmly grounded in the data.
The researchers frequently revisited interview transcripts
to assess the consistency of interpretations with
concrete statements of participants.

Finally, we acknowledge that
our work is not complete or conclusive
even though we claim theoretical saturation.
This reflects
the inherent unfinished
nature of qualitative constructivist inquiries.
From the very start we
recognize that theory construction is
context dependent and conditioned
by temporal, relational and cultural factors.
Instead of searching for universal truths,
we prioritize
the iterative co-creation of knowledge.
In accordance with this epistemological basis
we understand theory construction
as a continuous process
that goes beyond the publication of our work.

\subsection{Transferability -- External validity}
External validity and transferability
describes how well
results are applicable to varying contexts.
Qualitative research searches for a deep understanding of the particular.
Knowledge constructed from qualitative research is context dependent.
Therefore, we do not claim universal transferability
of our findings.
We instead provide a lens through which
the effect of phenomena similar to the ones
we investigated and present
can be investigated.
We provide this lens
by explicating how concrete conditions
influence our interviewees experience of software testing
and how those conditions feed into a
socio-technical dynamic which we describe in a theory.

Our study involved
software developers experienced
with cooperative software development
processes in well-established companies.
This means that
experience situated in vastly different
contexts such as start-ups or small open-source projects
are underrepresented in our data.
Additionally, the reliance on a single data collection method (interviews)
limits the variety of perspectives
that was included in the construction
of the theory.
Methods such as document analysis or participant observation
could have enriched the data in which our theory is grounded.
We mitigated this threat of a lack of variety of perspectives
by ensuring to interview developers from diverse roles
with varying levels of experience,
active in various industry sectors,
and working in various countries.
Diversity extends the potential applicability of the findings
to a broader audience within similar social and organizational contexts.

We address threats to the external
validity of our work by presenting
the results of a focused literature review
in \Cref{sec:form}.
We analyze what other scholars
write and compare their concepts and theories
with our work.
By juxtaposing our results
and the results
of other scholars from
various disciplines we do not only
identify new vantage points for
software engineering research,
we also demonstrate that our systematic approach
is able to independently reproduce results
even though it is dependent on context and co-creation.

\subsection{Threats to Groundedness}
With \textit{groundedness} we refer
to the extent to which the proposed theory
is rooted in the data
rather than being inspired by preconceived
notions of the authors.
We identify multiple threats to groundedness
and addressed each threat rigorously to mitigate
their effect on the credibility of our work.

First, preconceived ideas
can be a threat during the collection of data.
Interview questions and the way in which they are asked
can provoke answers that reflect preconceptions of the interviewer.
To mitigate this threat we
followed Charmaz' strategies for
semi-structured interviews
(i.e., intensive interviewing)~\citep[$\S3.1\P\text{7-10}$ p.58]{charmaz_constructing_2014}.
To avoid imposing our own ideas and language onto the interviewee
we asked for clarifications in interviews.
As explained in \Cref{sec:method} we used short nudges
(\interviewer{When you say big, what exactly do you mean by big software projects?})
and avoided imposing our ideas by asking open
and non-leading questions.

Second, preconceived ideas
can be a threat during the analysis of collected data.
To mitigate this,
we followed the systematic approach of reflective
and constant comparison.
Every interview was systematically compared with previously
analyzed data and used to refine emerging codes and categories.
We started the analysis of interviews without relying on
pre-established codes, remaining open to all possible theoretical
directions until we reach the end of the first stage of STGT.

Third, interpretation can lead to
a disconnect of results and raw data
rendering the construction of theory untraceable.
To mitigate this threat,
we maintained explicit links between raw data and concepts or
other theoretical constructs.
We provide extensive pertinent quotations
to demonstrate this linkage.
During data analysis we organized
links between codes, categories or memos and raw data
using a Computer Aided Qualitative Data Analysis (CAQDA)
tool to ensure traceability of theory construction.
We also
make the strength of evidence
for our conceptual outcomes
transparent by
declaring whether evidence
is conclusive, compelling or only suggestive.

\subsection{Researcher Bias}

As we explicate in \Cref{sec:method},
our work can only correlate our interpretation
of the experience of individuals with our own experience,
and the body of knowledge from the field
that is available and known to us.
We acknowledge that researcher and participants co-construct
meaning in interviews.
Managing and mitigating this bias is critical.
Instead of ignoring
or denying our positionality
as software engineering researchers
we integrate it into the research process.
In a registered report published before
we start data collection,
we explicate preconceived
assumptions in the form of
a list of hypotheses~\citep{swillus_deconstructing_2023}.
During data analysis we used
this list to reflect
in how far emerging findings were leaning
on those assumptions~\citep[$\S10.2.3$]{hoda_qualitative_2024}.
Following Charmaz recommendations
we also explicate preconceived notions
and positions in analytical memos~\citep[$\S7\P2$, $\S7.1.2\P6$]{charmaz_constructing_2014}
and reflect on interviews
and the construction of interview guides
before and after each interview.
During data analysis we use systematic
reflective methods to explore
and document biases and their influence
on theory construction.
By reflecting on our tacit knowledge,
relation to interviewees and
relation to the research subject
and including those reflections in
the process of theory construction
we mitigate the threat that our
preconceptions dictate
the development and outcome of our research.
\subsection{Ethical considerations}\label{sec:ethics}
We understand research ethics not as a set of hard principles and requirements
but as an ongoing discussion.
For the remainder of this section we discuss
how our considerations had an impact on
how we collect, process and analyze data
and on how we publish our results.

This study investigates and reports an analysis of
perspectives of human subjects on their lived experience.
We want to ensure that no participants are harmed through our study,
neither directly through the recruitment or data collection process,
nor indirectly through repercussions caused by the publication of our work.
We seek balance between transparency
and reduction of risk for all participants when
publishing our results.

To protect individuals from harm
we consider their right to privacy and self-determination.
For recruitment through the online platform \stackoverflow{}
we follow Nissenbaum's principles to protect contextual integrity~\citet{nissenbaum_contextual_2011}
and \citet{marwick_i_2011} to prevent context collapse \citep[see also][$\S\RN{5}\P2$ discussing context collapse in context of the present work in greater detail]{swillus_deconstructing_2023}.

Our study design was submitted to and approved by the privacy team and ethics council of TU Delft.
\section{Conclusion}\label{sec:conclusion}
This study set out to explore
why software developers decide
(not) to employ
systematic testing techniques
in software projects (\textbf{RQ3}).
We aimed to uncover factors
affect when testing is done (\textbf{RQ2}),
and investigated how testing related
opinions take shape (\textbf{RQ1}).
We explored our research questions
using socio-technical
grounded theory (STGT) and
constructed a theory
that explains how testing strategies emerge
from a non-linear, recursive process that
is influenced by technical,
social and organizational factors.

The systematic analysis of 19 semi-structured
interviews with software developers
revealed three categories
of conditions for testing.
Firstly, testing happens when
the testing infrastructure,
the application and business domain,
testing mandates,
available resources
and a project's vision
\textit{afford} it.
Secondly,
\textit{socio-technical aspects}
like the software development processes
employed by a project,
developers' concerns of
safety and responsibility
and the complexity of a project
condition when testing can be employed by teams.
Thirdly,
we argue that
dogmatic perspectives of developers
impact when testing is considered.
A project's testing culture
and perspectives of (online-) communities
influence how and when testing is considered.
In summary, developers test when
organizational-, social- and
technical conditions support developers to efficiently
pursue testing in a way that
is perceived as valuable for the project
and as worthwhile by the individual (\textbf{RQ2}).
\revised{Strong opinions are likely to be rooted in experiences
which feel profound or meaningful to developers (a posteriori).
Reflection of experiences within the social environment
in which software development is
practiced seem to shape opinions
more than individual preconditions (\textbf{RQ1}).}{2.7}

Supported by the categorization of
conditions of testing we establish
three novel testing related concepts
(i.e., \textit{testing signatures}, \textit{testing echoes}, and \textit{testing efficacy})
and construct our theory of emerging testing strategies (ETS).
Contributions in the form of testing artefacts
are enabled by reflective
(social) processes
which are in turn stimulated
by the presence of testing artefacts.
Testing strategies are not placed or
developed in a linear way but emerge from
recursive processes as
decisions to avoid or engage in testing
continuously alter the conditions
that necessitate those decisions (\textbf{RQ3}).

Implications of the theory we construct
are especially relevant for practitioners
who want to understand how and why
attempts to establish
testing practices succeed or fail.
The theory we present
makes organizational, social and technical
circumstances which impact testing practices
visible.
Used as a lens for analysis of specific environments,
it can be used to make possible
interventions for dysfunctional testing strategies visible.
Additionally, the theory identifies
new vantage points
for software engineering research:
\revised{it makes the connection
of technology creation and
the social and organizational environment.}{2.7}
Establishing this connection
prompts future work
to recognize their interplay.

While the study provides new valuable
insights into software testing practices,
it is important to consider the limitations
of this work.
The theory presented in this work
and the knowledge
on which this theory is based,
was co-created
by researcher and interviewees
and is therefore
bound to specifics.
In accordance with the constructivist stance
and the research methodology we choose (STGT)
the study embraces this dependence.
\revised{Instead of providing generalizable answers,
this study provides
a lens that can be used
to investigate the
composition and configuration of testing conditions
and their effect in software projects which
are not accessible to us.}{2.7}

By uncovering the socio-technical connection between
testing artefacts and collective reflection of practice,
this study not only advances
our theoretical understanding of software testing, it
also refines our understanding of
the experience of software developers.
It does so
by providing
a concrete catalogue
of conditions
for software testing,
and a theory that suggests their interplay.
Our study 
contributes to the groundwork for
investigations into software testing
which embrace testing not only as a technical
facet of software development,
but as an experience in which
human- and social aspects are entangled
with organizational and technical circumstances.
\section*{Acknowledgements}
This research is partly based on a registered report submitted and accepted at CHASE 2023.
Its scope was later revised and extended.
We acknowledge and thank the committee of CHASE 2023
for their evaluation and constructive feedback
which was integrated in this work~\citep{swillus_deconstructing_2023}.
This research was partially funded by the Dutch science foundation NWO through the Vici ``TestShift'' grant (No. VI.C.182.032).
The research was approved by the by the Human Research Ethics Committee (HREC)
of Delft University of Technology (TU Delft).
We provided a declaration of the purpose and scope of our research
to all participants prior to interviews.
The declaration clearly states how we register, analyse, store and publish data.
We explicitly asked for their consent to these conditions before each interview.
The opening statement of the interview and the consent statement can be found in a public data repository:
\texttt{\href{https://zenodo.org/record/18271664}{DOI:10.5281/zenodo.18271664}}.

\bibliographystyle{spbasic}
\bibliography{testing-tales}


\appendix
\section{Transcript excerpts}\label{appendix:interviews}
Interview excerpts presented on the following
pages exemplify the groundedness of our theory
in the perspective interviewees shared with us.
The excerpts represent only a selection of responses
to make the link between what was said and
our interpretation transparent.
All participants provided informed consent
to have excertps of their interviews quoted in our research output.
Questions and remarks made
by the interviewer are indicated
with a microphone
sign (\faIcon{microphone}).
All other passages, including those that are prefixed with a speech-bubble (\faIcon{comment})
originate from interviewees.
\revised{To protect our participants' right for self-determination,
names and identifying details have been removed.
Beyond that the original transcripts were only edited for brevity and clarity.
We indicate those changes with square brackets.
For convenience
we indicated the role, years of experience and industry sector of
interviewees
next to the interview identifier (e.g.,~\ref{subj:2})
of each numbered block of
interview excerpts.}{1.14}

\paragraphlf{\ref{subj:2} \--~Expert (10+) Finance}

\ievidence{%
I think the biggest
value of the project is
not its implementation.
It is its
test suite that we have build up over
all those years!
\interviewdetails{I think the project is 8 years old now.
And the simple fact that
it still runs is because we have all
those tests covering all these cases from
all these companies that use it.
They complain,
we fix and now
if you make a change,
and if you know all the test
parts you know you are still good to go.
Getting all that test coverage
and all the confidence that companies have
and all those millions of engineers that
are willing to bet on
this project that has been running for such a long time
is the value.
Its flawed in some sense.}
It could be better.
But if you have to choose between
a perfect tool that you don't know works
and a tool that has been reliable
and that has community knowledge and all that
you gonna choose the second one.}{2}{0.1}{221}

\ievidence{
When I merge a change I want to
be confident.
I don't check out code locally.
I literally do not.
I don't know
if I even have a clone
of the project
on my machine.
What I do
when I open a PR -
and the sort of change makes sense from a
conceptual perspective - I first look at
the test.
Did they change any?
If not then I know at least all of that is covered.
And if the CI is green - cool,
all the tests are running.
Then i look at the new tests,
now fixing
what didn't work before.
Or if there are test changes:
does it actually improve or is it
regression. And then
I start looking at the implementation
But I start with tests.}{2}{0.105}{262}

\ievidence{%
  \faIcon{microphone}: Do you also
consider a strategy
or something like this to tackle different things?
\faIcon{comment}: Yes. And
my opinion could be controversial.
I think unit tests
are overvalued.
In the sense that if they only test
one specific unit.
I do value unit tests in the sense
that they test one thing.
But what you see with the project
is that we do not actually write unit tests for
internal implementation.}{2}{0.11}{388}

\ievidence{%
\faIcon{microphone}: So when you say ``big''
what exactly do you mean by
big software project? Or
big company?
\faIcon{comment}: I would say
that the project has
- I mean, yeah, what is big?
A couple of
like 10 modules and a couple of hundred
s of classes.
\faIcon{microphone}: Interactions between\ldots?
\faIcon{comment}: Yes, it has at
least three components one of which needs
to talk to a database or a network,
one of which does business logic
and complicated stuff.
At least
a couple of dependencies where you need to interact.
These kind of things that always
complicate testing.}{2}{0.125}{00:22:23}

\ievidence{%
I don't think that university could do anything
different because if you give a person that
kind of complicated project -
if you want to teach them testing and
you shove this whole project
to them - they are not gonna
understand testing.
You have to start small,
so you are sort of in
incline.
I think the approach is good
but I am sometimes worried
that people then over-fit on those
things and then they
come into industry and then its like: ``what??''}{2}{0.13}{554}

\ievidence{%
I think
there wasn't direct
influence from people.
But you definitely noticed that
you work on a project where
there were people before you that defined
the way to test.
You sort of take that over
without explicitly listening
to the person.
You see
the influence of them
and I have done that myself when
I wrote documentation on how
you should do end-to-end tests and
the whole team started doing it.
They don't need to specifically know that it was me.
It is more like I define the
standard of how you should do it
and then you see the team
doing it because they also
made the mistakes themselves sometimes and
the approach works.}{2}{0.15}{653}

\ievidence{%
If I get 10 PRs [pull requests],
5 of which do unit testing in a small
sense and 5 that do it in a
larger sense, the 5 that
test in a larger sense - from a user perspective -
i am more easily to review and therefore
get the merge quicker.
And you get
the 5 small ones that just get stuck
there and they
become dreadful PRs
and that is when you
learn ``Oh''.
If you notice
this for yourself you can also
point people to it: ``if you do this
then I can easily review it.''}{2}{0.2}{687}

\ievidence{The one problem
with build systems is
that they are not deterministic.
It is just complexity all over
the place.
And everybody is contributing to it
so its not a stable system and then,
writing tests for an unstable system is
difficult whereas [the open source project] is a stable system.
You know that
the test stays.
But here, one month you
do this one thing and you might write a test for it
and then a year later you completely change it
because it doesn't work anymore.
And that's okay
but then you run those tests
and what did you gain out of them?
I mean after a year? Well, not a lot.
\faIcon{microphone}: Is
this sense of instability
something that makes testing hard?
\faIcon{comment}: Yes, but its
not just hard its also
that it doesn't make any sense in
that way.}{2}{0.225}{804}

\ievidence{%
Sometimes you just
work on a project
that for the business
size makes sense and then
nobody really
knows how to test it.
There might be
some approaches to test it but
nobody really knows.
But you still need to deliver
it for the business so
that is a
sort of priority problem then.
Like sure, you can invest a couple of months to
learn how to test it, or you just build it
and get the confidence in a different way.}{2}{0.25}{879}

\ievidence{%
It becomes more of a conversation: ``Okay,
do you consider this a requirement?
What would we need to change?''
And the conclusion might be:
``Oh, we can
change it and then it becomes testable''.
But it might also be that the conclusion
is ``no''. And that is okay.
Right? Ultimately you have to deliver
value.
And yeah, I
want to write tests but
sometimes I can't and you have to
be okay with that.}{2}{0.75}{912}

\ievidence{%
So I see a big
value in tests if you already have
tests. And this is a big
problem.
When tests exist
and people can
copy paste a test, tweak it
a bit for their use case,
fix the edge case
and continue with their life,
you see that you don't have
to ask engineers to write tests.
They will just do it because
they want the confidence, right?}{2}{0.5}{936}

\ievidence{%
Engineers will always want the confidence
and if you give them any
easy way to get that confidence they
will choose that.
It is just that sometimes
when you have no tests
they don't want to
spend all that effort getting that
first bit of confidence.
But the problem then
is you have to fix that
to get the snowball rolling.
And this is
mostly a business problem.
How do you get the
initial amount of tests in there so
that you get the ball rolling.
I mean its not a problem
just about tests.
Its also with loads of other
things but it is basically
a sort of culture change within a
project.
How can you build up
that initial suite of tests
so that it becomes so easy
that people would just naturally do it.
That is the ideal goal
but that is difficult.
That takes a lot of effort,
especially if the project was
build with no testing in mind.
And if you spend three weeks on writing
one test, that is a lot of effort
but it might mean that after
those three weeks the next tests takes you
three days.
And then, the next test takes you three
hours, then three minutes
and then you get what you want.
But it already
took you more than a month
to get to that position.}{2}{1}{949}

\ievidence{%
So in a sense,
that when you contribute to a project and everybody
knows it is so easy to write tests
that people feel comfortable to
say: ``Sorry, but you have to write a test''.
But there are projects
where everybody knows it is difficult to
write test and then people will
say: ``Okay. That's okay
because we know how difficult it is
to test''.
Then it becomes sort of a
culture within a project - there are projects here
that we know are
untestable.
So I am not gonna ask
you to do that because
if somebody would do that to me
I would say: ``Ah come on, really?
You ask me to invest three weeks to fix this?
Really, me??''.
Whereas in the
other projects its three minutes.
Like: ``Come on!''
Then I will be
okay with it.
So that is the culture within
a project.
As in: if I
know its easy
people will expect you to do it
but if everybody knows its difficult
people won't.
And that is the culture you build up.
This also means that if a project was
difficult to test and you made some changes
you now need to change the culture
so that you convince people: ``Nono, it is easy
now''.
Now you need to
actually expect people to do that.
But that takes time
because initially everybody is just used to say
``Yeah, it is fine. It wasn't t stable. Whatever.''}{2}{2}{992}

\ievidence{%
It happened a bunch
of times when I am having lunch with my colleague
and we have conversations about testing.
We do disagree on some topic and we just
have a conversation and sometimes,
like a couple of weeks ago,
we had lunch and we had
a whole discussion. I said:
``why do we have this?
can you explain it?''
And [they] couldn't.
But in that process
[they] realized. ``Oh, right''.}{2}{3}{1131}

\ievidence{%
I start asking questions and I
the ideal
scenario is that I come to a team
and I ask ``Why is it this way?''
And they say: ``We don't know''
and then it is like:
``Well, it is clearly a problem
now in this way. Can we figure
out what we need to do?'' And then you get
into discussion where they say:
``Yeah, actually we should have done this here''.
So I literally
had this Tuesday.
I had a meeting with
a team and we discussed
a particular problem.
And I just gave them
a bit of information.
It was a one hour meeting and after thirty
minutes of discussions
they just had a whole conversation
within their own team and I was just listening.
I didn't know what they were talking about because
I don't know anything about their domain
It was just ``Yeah, yeah, why is this here?''
like: ``We should have done that!''
And I am like: ``Amazing!''.
That is exactly
what I want. I want you to ask
questions about your code if its a problem
and then you are probably the best person to answer that.}{2}{4}{1215}

\paragraphlf{\ref{subj:4} \--~Expert (10+) Finance}

\ievidence{%
I was the one that went abroad
to see the project
running in production.
And after the first day,
after the software was running,
there was a big crash
that was very impactful for the business.
And that made me reflect on testing
a bit more.
I remember going back home
and thinking: You know, I learned all this
testing stuff at the university I never really do
it.
I know JUnit and I don't really use it.
Time to change!
And then I remember coding a project
myself, and run it, and it was in C\#.
Test-driven development.
So that is how I got into testing.
And I started to enjoy it}{4}{1}{00:06:25}

\ievidence{
If you get books on software testing,
they are very theoretical.
They explain the challenges of testing.
You learn basic techniques
like domain testing et cetra.
You learn
a small piece of specification or whatever.
And now you are in a
big project with thousands
and thousands lines of code.
How do you test it?
There is a big difference between
theoretical testing in one
small program example
and testing
in the real world,
where you also have pressure.
\faIcon{microphone}: Pressure in terms of
what exactly?
\faIcon{comment}: Pressure in the sense that
your goal there is to deliver software,
right?
And if you don't know how to do testing
then maybe it takes too much time for you to do it.
There is always, of course, the option to go
and say: I am not really ready to make this software
myself. I need to stop and
study and improve a bit.
But that never
happens. You just keep moving forward.
And your mind is really busy to
understand the business.
You want to deliver value,
you have lots of technical questions:
How am i gonna implement this?
How do I connect to a gas pump in that
200 MHz device, right?
Who cares about testing? I had other things in my mind.}{4}{1.002}{333}

\ievidence{I think testing can become really really hard if
if you are
in a software system that doesn't help you
to write tests,
right?
So you are in a software
system
where to write a test you have to
spend a lot of brain power.
``Okay, how do I have to setup
the data so that I
execute my test?
Now that I did this, how can I make the assertion?
How do I get the data back
from the system so that I know that the system behaves
as expected?''
If this is
very hard, what happens
if you are a very motivated developer - the person who
writes the first test - its gonna take you a lot.
The person who writes the second test
is gonna take less because the person
learns from the first
and does lots of copy and paste from the
first.
But the person who is going to do the third step
and maybe the fourth step might stop and say
``This is not productive.
I am gonna
just go and implement the next feature.''
So I think what happens a lot~-
in out of scale software systems~-
is that it is
hard to test
and because it is hard you are
gonna do it less, you know?
Because it hurts.
What you need to do there is, you know,
make sure that it is always easy to write tests.
Identify what makes it very hard
for you to write tests and create a
framework or create a small API,
a small utility method
that makes that job easy.}{4}{1.005}{596}

\ievidence{%
I started
to write tests and then I thought:
``Well,
it is taking me too long to
write one test''.
I knew that I would be lazy
and that I would not really test everything
so I stopped and I refactored
the code to make sure that I could
write tests in a much easier way.
Now, if you go to our test suite, 90\% of the test
methods start with the same line.
The same method call that is called ``run''.
Its a method that we created
and it calls the engine behind
the scenes
so that the developer~\--~the tester
doesn't have to do anything else.}{4}{1.01}{655}

\ievidence{%
The test would have to navigate through
fifty pages before going to the place
that you really wanted to test,
so my suggestion there was that
this is the hard part.
You are doing 50 steps
before you are getting to what you really
care about and then test.
Why don't
you make these 50 steps easier?
So then the goal - what we did - was,
we created a web API,
that would run together with the software
doing testing.
And this web API would offer
methods that would do those 50 steps
but much faster, behind the scenes,
right?
Just putting stuff into the database et cetera.
And in the beginning when I proposed this
to the team-lead they were like ``No, you
are crazy. That's gonna take lots of time
to do this. Hard to know if its
going to pay off.''
And then I said:
``I am here for ten days, lets
just invest three or four days on it
and see what happens''.
And then, in these three or four days
we were able to write a very stupid web
API for the basic stuff
and then developers could very quickly
start to write tests because those 50 steps
that would take a developer - i don't know -
2 hours of hassling around with
selenium and
selecting the right ID in HTML and so on.
This was fully removed.
And so they started to write tests
and then that payed off
and then they just spend the rest of the time
writing those things.
And then
the default was changed to ``Okay, we
know we have to create this, then
how do we create this
in a maintainable way, in a scalable way''
So we switched the problem.}{4}{1.2}{828}

\ievidence{%
I think this is one of those things
that people express in the community.
And then it becomes a rule.
At some point people would just say:
``you should never ever write
code or change your production code because of
testing'' or
``you should never write production code
that helps testing'' and so on.
There was some sort of discussion
there: you know should
I change my production code to make testing easy?
And to me it was always the other way around.
It was always a trade off.
I would change the way
how I code my production code a little bit.
Maybe I am gonna add some extra complexity there,
but that would pay off because
I am going to test it in a much easier way.
And I feel that that is usually the barrier that I see
even here.
At times
I talk to engineers and they are like
``you are telling me to change
my production code test??'' and I am like
``yeah, are you testing right now?''
``No''.
``Exactly!''
I feel this is a big
thing that we need to change in the community.
That you should invest on your testing infrastructure
maybe as much as
you invest in your production infrastructure.}{4}{1.5}{894}

\ievidence{%
I think you have to see testing
as important
as production code and if you do this
then, when you have to take a decision
to make testing easy,
you are gonna do it because testing is
as important as production.
I think it is all about
putting things at the same level of importance.
And I understand if for you production is way more
important then testing.
Whatever decision
hurts production, you gonna bash on
this decision, right?
This is something that I hear a lot
when trying to convince people.
Because I am not telling
you ``Just create this private
method here in you test suite, that will take 10 seconds'', right?
I am telling you that you have to invest
X weeks of work.
And then it is almost a project,
right? And then people are more concerned.
``Should I spend this amount of money
on my testing?''
\faIcon{microphone} Because they never experienced
how well it can work?
\faIcon{comment}: I think
its because we are not very used to
large software engineering and testing projects,
you know.
You see companies
now, having
dedicated groups of engineers
working on platform engineering.
And part of the
group is working on testing.
Like a dedicated group working on it.
And usually, product teams
are not used to investing
hours and hours
and thousands of euros in testing.
That's something to change.
And I think if you do this from the very
beginning you are going to save money because
you gonna test more, you gonna test more from the very
beginning and have less bugs.
I think that is - socially speaking that is
a challenge. Its
something we need to switch in peoples mind.}{4}{1.6}{929}

\ievidence{%
When you put in a little extra
energy to make something better.
You know, ``This is done, let me
improve it a little bit to make it a little bit more
beautiful''.
You don't
take this to the extreme,
but you have this thing in you
that doesn't allow you to
submit a horrible merge request,
you know?
You write the code
and before submitting the code
you do a small refactoring to improve
the legibility of the code.
I like when the person
is okay with
taking 10 extra minutes
to make what she did a little bit
better.
Because
if everyone is like this,
then development gets so much easier.
The green garden story.
If everyone is there to make sure the garden is always
green, then
it just becomes so easy to develop.}{4}{1.7}{1140}

\ievidence{%
You know we had
the chance to do it from scratch.
No legacy.
So we wrote it with all the best
practices and there are lots of tests
and classes are small and so on,
and then he opened a feature
there and he submitted
the merge request and the merge request is full of
tests and then i said to him: ``Can you imagine if all developers
would do it like this?''
And he said
to me: ``Its much easier to do this here because everyone
does it like this''}{4}{1.8}{1188}

\ievidence{It was nice to remember stories from the
old places. It has been a while since I
told them. I had a lot of fun in this company.
At some
point the consultancy became
popular and I was basically
getting on an airplane every month and going somewhere.
I was always visiting different companies.}{4}{2}{00:46:07}

\paragraphlf{\ref{subj:6} \--~Early in Career (1-5) Finance}

\ievidence{If you are
a very large company
like [company name] then
there is no way around it [testing].
And if you are in a smaller company
then I think the excuse would be
that you are with very little people so you can
manage it [software development tasks] more easily.
In large companies you have [testing] teams
of course and [name of fintech company]
might be even more to the extreme
because it is a financial company.
Even a bank.
So there are even more regulations etcetra.}{6}{2}{00:19:25}

\ievidence{%
\faIcon{microphone}: You also mentioned
this aspect of your work [testing]
changed in comparison to
before?
\faIcon{comment} yeah, for sure.
\faIcon{microphone}: How?
\faIcon{comment}: In the sense that its more
common or - I don't know
if there are any guidelines but
it is more common at least or
its assumed that you test your stuff.
At least unit-test.}{6}{2.5}{608}

\ievidence{It doesn't really make sense
to write integration tests right now because
we can't insert the
the unprocessed entities
because they originate from
another database or table
through triggers and its a whole complicated
process. Its also something that I had to
of course learn.
When you join a new team you need to know
all the business logic of it.
At least I like to know because then I have some
sense of context.}{6}{3}{00:28:29}

\ievidence{%
Okay I want to set up a good proper
structure so that we can test everything
or not necessarily every line but
every major feature et cetera
but because the application was very big and
the features were so different
and they were already very old
then yeah that was just
too much so then that kind of like
got de-prioritized because
I was thinking too far
or too big right?}{6}{4}{1575}

\paragraphlf{\ref{subj:8} \--~Early in Career (1-5) Finance}

\ievidence{I don't know what the normal structure is
but I guess it would be similar to ours
so we go through a whole
first business process of
coming up with new things and
then
they are being introduced to us
and then we implement
those, we test it first
locally and it needs to
get accepted by certain amount
of people in the team
and then we go for a release
process which is
just a bunch of testing
and checking for mistakes
and different types of errors
and then
its being
tested on a bigger scale rather
than just
from our team}{8}{0.15}{363}

\ievidence{%
That was definitely my fear
when I started working and my manager was
amazing in comforting me
and telling me how impossible
it is for me to actually
mess things up in the code because
it is just impossible
like you said.
There are all those tests
you need to go
through and you need to write.
And to merge the code
it needs to pass the whole pipeline
of all the other tests.
It needs to be checked by my manager,
by another
engineer and then
even then its just like on
another branch.
Its not out there
in the public.
It still has to go through a bunch of
different tests.
And I think that was very comforting
as a beginner.
That there is
a whole\ldots
The way that this whole infrastructure
is built
it prevents - of course
mistakes happen -
but it prevents
most of them from happening}{8}{0.2}{00:58:09}.

\ievidence{%
\faIcon{microphone}: You said its too complex?
To test?
\faIcon{comment}: I don't know. I did say that because
I was just thinking about some of\ldots
I just can't
imagine. If if someone was not testing
a software in the beginning
and then you have this
big complex thing that
has been going on for years
and then how do you start?}{8}{0.5}{1048}

 \ievidence{%
So if I am just hard coding
a line saying ``print
hello'',
this will work no matter what.
  If i need to get the "hello'
from a certain API
and the connection might be lost
there is just something that might go wrong
then I
want to test it.
So anything
that is
dependent on something else,
I guess, is worth testing
because It might go wrong.
Anything that is very important,
 - even if it seems
easy - and should always work
is worth testing
because if it is very important
then it is just better to play it safe}{8}{0.7}{1149}

\ievidence{It is usually like really
two lines
code so
it never feels like an effort to
add those tests.
\faIcon{microphone}: Do you think you also benefit
from the work that your colleagues put into it?
\faIcon{comment}: Oh for sure. I mean
the tests are very repetitive
so it is mostly
the same test over
and over but just checking different things}{8}{1}{1197}

\paragraphlf{\ref{subj:10} \--~Early in Career (1-5) Finance}

\ievidence{At some point
\textbf{you} will be responsible for maintaining it.
It is that fear-factor, that
the maintenance needs to be done
and it needs to be done
efficiently}{10}{0.25}{00:29:19}

\ievidence{If you
get past that state
that you are going to build some product that
has the goal to be released,
then I think you
won't be able to not
test your code because then at some point
somebody would say ``hey this can't go live''.
Because you have not proven this, this and this.
\faIcon{microphone}: Because of lets say policy?
\faIcon{comment}: Yes, exactly
\faIcon{microphone}: You have to do it?
\faIcon{comment}: It is not a choice. You have to do it.
So you might as well start directly.}{10}{0.3}{486}

\ievidence{It was really focused
on results. For example,
you had to
model some kind of flow
and you would model the flow
and then you would say: This is it.
This is the end of it.
So I did not write a single test for that entire
program as well.
\faIcon{microphone}: You did not see the necessity
to do it I guess?
\faIcon{comment}: No, and then \--~
yeah, I think that
changed a little. A lot I think
during the university
courses that followed
and eventually even when writing
a full-fledged application. I think it also
becomes really apparent when
you meet a certain
threshold of lines of code
because then your overview is gone
If you have a programm with
150 lines then you are able to test it kind of
input output based
or something and you have seen
multiple runs
but at some points things are connected in a way that you can't
see anymore and then if you
make a change it might break some other part
of your program and that is really when the horror starts
because then you have no clue if
a change you make will not
add like a regression
or something like that}{10}{0.5}{00:33:17}

\ievidence{%
They can provide a full
\textit{.jar} \--~ that is your
code base that you are going to run \--~and you can
simulate exactly that and
test it end-to-end.
\faIcon{microphone}: So
the testing tooling that you
get from the project provides
all you need and it even encourages
a certain way of testing?
\faIcon{comment}: Yeah.
So, in the team that is one way to test
and that is the
fastest way to test but even
in that team, one of the team members
has written an entire
compose-way of running an integration test.
It will actually simulate as if it
is running on a container
so we also have integration tests.
If you want to merge to main its
a directory with 20 tests and those
tests know business logic we
have documented and because it is documented
it is also associated to an
integration test that tests for
us that no changes that have been made block
or break the functionality that we offer
to customers.
\faIcon{microphone}: So apart from the
thing that you get, you also have in-house tooling
to make it even easier.
\faIcon{comment}: Yeah but also to
proof that
it works, that its also
proves to work for a customer.
Not only the code flow but also \--~yeah,
what if something else happens or
the container in which
it runs has issues?
We really simulate the entire environment.}{10}{0.7}{990}

\ievidence{If it is only for myself
and it is a small
project, then the risk of when it
goes wrong maybe that is also one to take into account
the impact is very low. namely
I can't download the package I have to do it the old-fashioned
way then
if those three converge then
I would be okay with}{10}{0.8}{1132}

\ievidence{%
I used to push like: ``hey this is a pull request'''
and it had like 1500 lines of code
or something.
So yeah,
what can you do with such a pull request?
At the time there was also
no push back to push PRs like that
and here,
some people have to actually approve my pull request
and even more than one person: the four eyes principle.
If I am going to
push 1500 lines they are going to
laugh in my face and say:
``do you have a
good reason that this is a necessity?''
But for most of the PRs that is not the case.
Again, it is also like a
trial and error kind of
way and I think also at that smaller
company I worked for it
worked a little bit like that.
That you would figure out what worked and that would
stick and then you would continue to do that.
It wasn't an agile based
company but at the end,
in 2021 we started having
stand-ups and stuff like that
where it used to be only ad hoc
et cetera and every single
time we would learn from
what we could improve.
I think that is also a very nice environment
to be in. Like ``hey
this is not working with such hard PRs!''.
\faIcon{microphone}: So the nice environment also comes from the
approach and attitude of the collaborators
not only from the
nice tools.
\faIcon{comment}: Yeah. both.
I think the human factor is in the sense
that you are willing to
grow I guess as a developer
so if you see a benefit and
you are open to it or if you see an opportunity and you are open to it
and they are okay
with implementing it even though it might change
how you used to work
then it might be a complete mind
switch that you suddenly
have to write code again}{10}{0.85}{1480}

\ievidence{I think it was nice
to tread down the memory lane.
A reflection of
all I actually did in
a linear way.}{10}{1}{01:00:17}

\paragraphlf{\ref{subj:1} \--~Senior (10+) Software}

\ievidence{%
You know, testing not my favorite topic.
But we can talk about it.
I mean, I've had a lot of experience,
especially in corporate with testing in private with lack of testing, maybe.
But yeah, I mean, we can talk about any aspect of that.
\faIcon{microphone}: Yeah, why is it not your favorite topic?
\faIcon{comment}: So at my former employer, for instance, QA, you know,
like testing and all that stuff is - essentially, it falls on the developer to write tests.
And then, they're constantly like - and this is not just me, this is everybody -
they're like, [imitating annoying voice:] ``Hey, you know, you should really write some end to end tests for this''.
And everybody's like, [imitating annoyed person:] ``Yeah'' you know, and
Some people are really good citizens,
and they do some great tests
and then the code changes and the tests break,
and then they have to maintain those tests.
So basically the more you did,
the more you contributed positively
to the test --- to the code base
in terms of testing --- the more
responsibility you have later, right?}{1}{1}{00:05:17}

\ievidence{A fundamental problem with testing is,
  it's kind of an external thing that you have to do, right?
  \faIcon{microphone}: Yeah, but but you also mentioned,
  if you want to be a good citizen, some people take it up on themselves and do it.
  Or what do you mean by, you know, want to be a good citizen?
  \faIcon{comment}: Yeah, so well,
  this is not about my personal stuff, right?
  This is about about, you know, my corporate experience.
  So in corporate, you know, there's always somebody watching in a sense.
  But then, you know, there are always degrees where you can shirk.
  You can avoid doing the thing, right?
  So, in that sense, the most responsible people are like:
  ``I'm, really working on testing''.
  But then for me, because testing became kind of an obsession, when I was at a corporation,
  actually.
  And then it became like:
  ``hey, now you're getting paid to do testing, or you're getting paid to….''
  So, you know, in a sense, the good citizens thing, it does change.
  You know, the people who are like, really upstanding people
  will probably write more tests.
  They'll probably do better tests, they'll probably take it more seriously.
  But in general, in corporate, everything shifts to be a prime.
  From being a secondary motivation to being a primary motivation. Right?}{1}{1.5}{69}

\ievidence{What's exciting for software developers
  is shipping something to people that use your code.
  So there's a motivational problem,
  there's a maintenance problem,
  there's, there's just a time problem,
  like, you know, your boss is looking at you like:
  ``Hey, you know, what did you do?''
  And, you know: ``Oh, by the way, I forgot. Did you write tests for this?''
  You end up with either, you know,
  very minimal tests or this or that,
  but I mean, it's part of the problem with engineering.
  It's why we build bridges that fall down build
  buildings that fall down.
  Engineers want to get shit done fast,
  you know, and there are often external pressures, too.
  But testing is, yeah, it's, it's not, it's not the most exciting topic,
  unless you're a QA engineer, in which case it's the most exciting topic.
  \faIcon{microphone}: So it sounds like a burden. When you describe it like that.
  \faIcon{comment}: I would say, for me, if I had the bandwidth and the time,
  and you were like: ``Hey, we're gonna pay you to do this, or you're a QA automation engineer, do it!''
  It's fine. I'm indifferent as to what code I'm working on,
  as long as I'm working on code.
  But as my own work is structured, as my corporate work was structured:
  Yeah, it's absolutely a burden.}{1}{2}{64}

\ievidence{It was one of these things where you're like,
  \textit{click, click, click, click, click,} put in a name, \textit{click}, put in a name, \textit{click}, put in a name, now get to the next screen.
    ``Oh, s**t, it's broken again! I'll fix it'', you know?
    You've done that eight times and even if you've done that five times, you're like: ``okay, let me see how to do it [automating the test case]''.
    It wasn't automation for safety, or for real end-to-end formal tests that I was going to ship.
    It was more like, how can I just avoid doing this?
    Like, testing this work as I'm developing.
  So it was much more like a developer tool}{1}{2.25}{00:12:42}.

\ievidence{%
In a sense, software development is a very iterative process.
And luckily, it's made for that, right?
It's not like building a bridge, which is not iterative.
Tests are nice. Tests, you know, really give you some sense of solidity.
However, when you get into it, when you get into the nuance, even a little bit -
tests are never complete.
You never test all code paths,
you never test even most code paths, right?
So you're sort of adding some sort of baseline testing,
which is really nice.
But you're also adding some - I don't know if it's proper to call it technical debt -
but you're adding some future technical debt for those tests.
You know, the work now increases, you're going to have regressions,
you're going to have regressions on the test only also,
and you're going to have regressions on the test.
You're going to have regressions on the testing framework.
Oh, selenium changed this. What the f***}{1}{2.5}{94}

\ievidence{I have a colleague and he's a staff engineer. And he is young. So he's really good. And he's like:
``look, all this testing is bullshit. Like even unit tests, it's all bullshit.
Unit tests have never prevented a bug from shipping. It's all like just a waste of time.''
So, he's not totally wrong, by the way. But it's like, kind of a nice, polemic position.
And it really serves like a social function.
He thinks that the people who spend like millions of hours on tests
are idiots, right?
So this is one of the things that I know you don't expect this answer, but it's a polarizing thing, right?
How important is testing to you?
Is testing a formality?
Or is testing like a fundamentally important thing?
And it probably maps to political leanings, social leanings and everything.
Your risk. How likely you are to speed, to break laws on the highway, you know?}{1}{3}{115}

\paragraphlf{\ref{subj:3} \--~Senior (10+) Finance}

\ievidence{%
Mandates where the employer is saying I need complete code coverage.
There's an element of wishful thinking there
because unless the developers also want to do this
unless the developers are fully engaged with the desire
to generate it - Why would you want full coverage?
It's because you want your software to be fully tested
Coverage is just a proxy for yeah to what extent do your tests?}{3}{0.5}{115}

\ievidence{I've been asked to write a program
calculation therefore my deliverable unit is
It's that that code the thing that does that the thing that I've been asked to do
whereas, you know an alternative looking at it is that's just one step in a
In a sequence of iterations
Is going to have to deliver
And reason why we have tests is because it allows us to make subsequent iterations because it means
you're able to lock in the
Behavior of previous iterations and then you can build a subsequent step
Knowing that you can compromise the previous steps, but I don't think that's possible with it with a complicated program
unless you have some kind of testing strategy or
Unless you you have some kind of guarantee that the same developer is going to be
working on that code
or that problem, the entire time
Because as soon
as you have somebody leave the organization
it becomes impossible to know what the purpose of
a thing is.
Generally things aren't commented,
aren't well documented or anything like that.
The tests are the only real guarantee of what a
systems desired behavior or systems behavior ought to have been.}{3}{1}{00:18:24}

\ievidence{So ``done'' includes that there's a test
With coverage covers all of the changes you just added.
The thing isn't done until you have that.
Yeah, and I think that sends a signal to the developers,
which is that actually
You do have time to deliver this
There's no such thing as being too busy to write tests
because tests is now is writing tests is part of the specification
The management has asked you to do and you know management
are the people who pay the bills
so therefore they have a perfect right to ask for this.}{3}{2}{261}

\ievidence{There's a famous sort of demonstration
on YouTube of somebody
showing that you can write decent software [using software testing practices] whilst drunk.
There are really good reasons why we do this
and yet, people tend not to.}{3}{3}{00:31:03}

\ievidence{the individual who might
  be feeling secure in his [developer-] role gets job security
  by having a code base that is so mysterious that only he can understand it}{3}{4}{00:32:22}.

\paragraphlf{\ref{subj:5} \--~Senior (10+) Software}

\ievidence{\faIcon{microphone}: I'm super intrigued by the ``spectacular failure'' you mentioned. What was going on there if you don't mind
sharing it?
\faIcon{comment}: Okay. So we did simple website stuff, but one
of the clients was requesting for a system to track judicial cases, I think.
It was a very special case of hiring partial hourly staff.
So it was very, very simple, but for us it was super complicated.
We never did software development at that scale before.
We treated it just as a web page project and there was like a two-page requirement spec that
we got from them.
We presented our solution after four months and it was very, very crude.
And they were very underwhelmed, let's say.
And the project moved on
like for a year.
We never delivered.
It was very, very bad, very bad communication with
everyone.
Everyone was mad to each other. Yeah, no, that was horrible.}{5}{0}{26}

\ievidence{Yeah, so after that initial work that I did, I left the company very,
very mad at software development in general, and I left with the feeling that there needs
to be a better way to do software development. I was pretty sure that there were like 20
years of experience in the industry. So I was really frustrated with the university
and what they taught me. I felt that that wasn't enough because they were reciting the
same things that we did wrong as the way to do things. So I reached into the general community.
That's probably the first moment that I started caring about community on technology.}{5}{5}{42}

\ievidence{I discovered agile
development and unit testing and that opened my mind completely. So in the first project that I took
after my first job, I started implementing those things and learning from my own perspective. I
luckily had a lot of leeway. I always worked on those first projects as a sole developer.
So I had all the tools that might be possible to implement whatever I needed.}{5}{1}{51}

\ievidence{The first project that I took
after my first job, I started implementing those things and learning from my own perspective. I
hopefully [sic: luckily] had a lot of leeway. I always worked on those first projects as a sole developer.
So I had all the tools that might be possible to implement whatever I needed.
And that's when I started doing unit testing from blog posts and papers and stuff,
and design patterns.}{5}{1.5}{52}

\ievidence{Discovering techniques,
  understanding how people worked, understanding the limitations of the
frameworks that I was using.
At that time, it was not very user friendly to use unit testing.
They started working on that later on. So yeah.
\faIcon{microphone}: Can you remember people, maybe from the community, or around you that influenced you
back then? Or was it really just you, the papers and your curiosity?
\faIcon{comment}: Yeah, I didn't have contact with people. I only knew names like Robert Martin or
all the authors that are now commonplace.
Not because I had the books, I didn't. I didn't
have access to those books. I just read the blog posts
that they did. And blog posts had this thing that they come from a place of
their understanding, but I didn't have the same background.
So I had to reconstruct that
background for myself.}{5}{2}{00:09:54}

\ievidence{At the beginning, I was testing the waters,
understanding what it meant for me.
If that [testing] was a viable solution.
Now, I strongly believe it's
a practice that is needed on software development.
Yeah, we build features.
But most of our work outside
code is probably more important than the actual coding.
Testing is not only for ensuring quality, but to ensure understanding. And I think
that for me is critical. It's part of my design process.}{5}{2.5}{94}

\ievidence{%
I believe that tools or practices are tools that you need to have in your toolbox and use them
when available. Testing is one of those things that like a pencil that you're going to use
every time. But you're going to use them in different measures, in different ways,
depending on what industry or what type of team you're working with.
\faIcon{microphone}: Yeah, so you introduce variations depending on the context.
\faIcon{comment}: Yeah, and like everything for context, for taking decisions you need context both on
the situation you have and the practice that you understand.}{5}{3}{176}

\ievidence{%
The level of testing or
other project management strategies~\--~you're using Scrum,
you're using Lean~\--~ you need to adjust the team to those frameworks.
And I think it should be the other way around.
I think that the team needs to find their framework. How to work.
So, every software developer is an individual,
but I also strongly believe that every team is an
individual and team composition is immutable.
If you change one part of the team, the team changes
completely.
}{5}{4}{00:31:43}

\paragraphlf{\ref{subj:7} \--~Senior (5-10) Software}

\ievidence{So testing, it's less relevant in my field
field specifically because
most of the time when I work,
it's a bit of an
iterative process and,
most of the time,
the end result that actually goes into
production is mostly just a machine learning model,
maybe wrapped up in an API.
I'm not even responsible for, for developing the API itself.
So the parts of code that goes into production on my side,
is often just
a ".bim" file that contains the model itself.
The parts of code that go into production on their [my] side
are being tested in a different manner and not like unit tests[sic].}{7}{0.5}{00:09:34}

\ievidence{%
So I think it's healthy to, to, to write tests and, uh, I encourage others [to write tests] too, like the project
I'm working on right now.
We also have like a front end development team and they do not have unit tests and we
have issues because of that, but, uh, they're pushing for a deadline right now.
And it's kind of in a backlog.
So once we get this deadline,
we can start working on unit tests and stuff like that.}{7}{1}{208}

\paragraphlf{\ref{subj:43} \--~Manager (10+) Software}

\ievidence{%
I encourage the teams
to see testing
or quality control as
their responsibility\ldots
I think it is the responsibility
of the software engineer.
So whenever I write code I write tests.}{43}{0.5}{474}

\ievidence{%
We were building a solution for
[a bank] and we had to make sure that the quality
was there of course.
Or that the software did what it should do.
And it [unit testing] also speeds up development
right?
When you can just
rerun your tests.
So there was a
framework
that helped to run
the tests and to verify that
things worked as they should.
So from the moment I started
using unit tests
I always
liked this
because I think it
helps me speed up
because I can verify that it works
if I refactor something.
I can check if it still does what it
needs to do.
\faIcon{microphone}: So it was a good experience?
\faIcon{comment}: Yeah, for sure.
\faIcon{microphone}: And
you said "we"?
\faIcon{comment}:Yeah there was a
small group of people.
We were
with 3 or 4
doing coding I think.
We were
looking at what
we thought would be good practice
to check for quality.
And, yeah...
\faIcon{microphone}: You said
you encourage people to do it nowadays
so you also
elaborated this with your colleagues back then.
You learned your lessons and now you
you encourage people
to follow that.
\faIcon{comment} Yeah, but to be fair I mean
I may have read 200 books on
software development and
many other things and
architecture and
well, there is a difference to
doing it because you think it is
required or that you think it is really
worthwhile to do it.}{43}{1}{538}

\ievidence{Well, I think
it should be an instrument
it should help you to deliver
better quality faster
so what you also
see is that it becomes a religion
and I have also seen that like
yes we need to make unit tests
they need to cover everything
so we write them
but did you think of the edge case
so do they add
value to what you are doing
and if your
software is
if the structure isn't right
then any change you do, you
have more work changing all
of your unit tests then you have
work in actually changing your
software so it starts to hinder
you
and I have seen
that as well and that is something
you should not
have so then you need to reduce the
number of unit tests you have
focus on the once that really add
value or you
need to improve the structure of your software}{43}{1.5}{605}

\ievidence{%
Well, it is documentation, right?
So testing documents
a part of the intent
you could say.
Looking at tests
you can understand more about
how the code
base was intended to work
what cases were considered.
So I also see
them as documentation
\faIcon{microphone}: They communicate
intents?
\faIcon{comment}: In varying degrees of
success
I would say [laughing].
But they can communicate
intent. yes.}{43}{2}{649}

\ievidence{%
I think I will probably have a more
positive view of somebody
but I can relate to that if
they do include tests and you see that
they are thorough
in what things
they consider because you do get more
trust in the mechanism so
I don't think that is
explicit for me but it might
make it easier for me to
sort of box it off. To say okay
I trust the outsides of this
I don't have to look inside and I can just
use it
}{43}{3}{691}

\paragraphlf{\ref{subj:45} \--~Manager (10+) Software}

\ievidence{%
If you are lucky
you have a release cycle of two weeks
and then you need to do some manual testing.
A lot of people
will be tempted to say:
``yeah its two weeks from now.
I can do it tomorrow.
I can do it monday.''
We need to get rid of that.
Actually we got that down.
We already reduced it
by one week by simply saying:
``We are going to reduce it by one week
you need to start testing on day one.''
\faIcon{microphone}: So how exactly did you do that?
\faIcon{comment}: This in particular wasn't too
complicated.
It was a matter of communicating
the intent.
Sending
various messages.
First, talking to the other engineer managers,
preparing them a bit,
getting a slightly
ever growing group of people involved
and getting them
on board with the idea.
And just communicating it,
giving everybody the confidence also that well,
there is a chance that
it won't work the first time.
But that is okay.
We'll learn from it.
Also, if because of this they discover that they can't actually do
it in a week, they will get the time to
fix their systems
in such a way that they can do it.
Gain confidence and
then we just did it.
That turned out fine.
}{45}{0.5}{359}

\ievidence{%
Well testing
the product
for the most part is automated
but part of it is not.
Doing those tests is not fun.
Nobody likes it.
Almost no one.
So thats
not an incentive.
Thats difficult indeed and
i don't think we fully
solved this.
Thats difficult in an organization
with 140 people but
of course there are a few things you can do.
One of them is
showing off examples
that might
motivate or inspire.
There are a few teams
that have fully automated their tests
and they don't have any trouble during
the release. So we try to make that
visible and tell people: Look
this could be you [laughs].
Its hard to get there but
you can. It has been done.
Another way is
obviously improving the
developer experience.
If you look
at any
persons job
there is one part that is really core
to the job. Your main specialization
where you are good at
what you probably, hopefully enjoy
doing. And there is
the other part
that simply has to be done
to keep the
world running. If you are a
PhD student that might be
the research. Its what you
love but you also have to teach.
Some people love it. You might love it,
I don't know but
there is a bit of a conflict
between them.
What I try as much as possible is
to give people the
experience, the developer experience
that they enjoy it
to a large extend.
That means
trying to make that
part smaller and smaller
and giving them the best tools they
can get so that they can work more
productively and don't have to
wait every time and
do a lot of manual stuff.
\faIcon{microphone}: You mentioned providing the infrastructure
\faIcon{comment}: Providing the infrastructure,
and removing the overhead.
Reducing the
time people need to spend looking
up details or "How did this
process work again?"}{45}{1}{387}

\paragraphlf{\ref{subj:47} \--~Manager (10+) Software}

\ievidence{%
There's one thing where every software developer communicates. It's
non-verbal, it's literal.
And that's the source code itself.
I mean there you can see
a number of different things that reflect out of that.
So when you make any change to the source
code - let's say that these days it is common that you make a merge request
when you make changes - you reflect on that.
You communicate why you make changes.
You usually
have some form of definition linked to that [merge request],
where you say this is why we're doing that.
But yeah, the big part of communication you're doing here as
well is actually the software change that you made.
All other things that you're doing are to
a large part next to that.
If you look at that in a very puristic manner
that means that the more
you have to communicate why you made that change and what you did,
the worse of a job you actually
did in the source code.
It's not really explaining itself.
There are even people that feel
that source code should explain itself completely and be without
documentation.}{47}{1}{258}

\ievidence{%
I think form and function do follow
each other.
I think there are quite popular effects
where people talk about the form of your team also changing based on the thing they're
solving.
So that basically architecture and form forms each other.
The architectural, the archetypal example of saying:
we asked a number of teams
to build a parser or three teams building a parser and it turned out to be a three-phase
parser, because they they divided the work in there parts so the architecture
formed in that way.
Yeah so that's basically kind of what you will get so
I would say that you're making me think - because it's an interesting proposal itself -
that you say that the form in which you have a team, like a team that is more trusting
that allows for more mistakes or kind of is open to
listening to the feedback of others has  an impact in the form of the software.
I think it
definitely has an impact on the quality of their software.
If it has an ability
even further than that i think that's very interesting.
I don't know. I'm going to think a bit about that.
That might actually be a very interesting uh reflection there.
So maybe you've stumbled on something like Conway's law but
instead for architecture, for software development}{47}{2}{375}

\paragraphlf{\ref{subj:49} \--~Manager (10+) Transport}

\ievidence{%
So I can't demand,
but what I will do is tell everyone about the cool TDD course
we're doing and advise them to come.
And if they don't come,
I will still like apply whatever I've learned to my code reviews.
And I will tell them like this unit is testing internals
while it should be testing user interaction.
So maybe we should do it differently.
\faIcon{microphone}: Right, so you encourage it rather than demanding it.
\faIcon{comment}: Yes}{49}{1}{576}

\ievidence{%
  I think it would have been really hard to improve testing.
If you have this whole bunch of untested code,
I don't really know how you get to a point where you have a whole bunch of tested code.
That's an investment that has a lot of business value on the long term,
but not a lot of business value on the short term, right?
And that wasn't really the thing that they were focusing on at the time.}{49}{2}{834}

\paragraphlf{\ref{subj:51} \--~Senior (10+) IT services}

\ievidence{%
It's tricky for us because I think everyone knows and agrees that it's
good to have lots of tests because we've all seen in practice that it helps to prevent
problems. It helps to prevent them early before they become big problems. So it's something
that's obviously good but it's also something that takes lots of resources to add. So there's
a continuous struggle to weigh this [investment in testing resources]
to balance this to see how much can we spend on tests. Also, even though people
know that it's better, there's also the culture to maybe not finish things to 100\% but to
have it working and then focus on other stuff that's more burning and often tests come a bit
too late on the priority list of things within a project. So then there are no tests or just
not enough to cover what we really need to cover.
\faIcon{microphone}: With culture, just so that I understand it right, what exactly do you mean with culture
in that context?
\faIcon{comment}: Well, if someone adds a feature or a whole sub-project that does not have full testing,
they will often not be pointed out and say, hey, you also have to add some tests. Sometimes
people might say this, but this is not standard. It will often be overlooked.
\faIcon{microphone} So culture then means the way in which collaboration happens and the way in which people\ldots
\faIcon{comment}: It's also a shared implicit understanding of if it's necessary or not. It may seem more
as a luxury while it would be better if it were a basic requirement.}{51}{1}{130}

\ievidence{%
\faIcon{microphone}: What would you say you would need to do it effectively?
\faIcon{comment}: So it's very hard not to answer more resources, but that's not the best answer because...
\faIcon{microphone}: Why is it not the best answer?
\faIcon{comment}: Well, because I guess any IT company, or maybe not, but at least anyone even remotely like
the one I work, there will always be a resource constraint. There's always 20 times more work
than you can handle. So it would be nice to have some extra resources to add testing,
but even if we had more resources, chances are that we would spend them on other stuff
instead of just the testing. So actually we should probably change something about that
company culture or have a different focus as a company. Not even to radically alter
the fully test-driven or something,
but just to change that balance that I mentioned earlier.}{51}{2}{147}

\ievidence{Yeah, the complexity of the software environment.
So how many different external components there are that need to be present to run the tests.
Of course there are techniques to deal with that, like make mockups of individual components.
But yeah, that makes it so complex that at least for smaller projects
and the high resource constraints we have, it's unlikely to be done.}{51}{3}{209}

\ievidence{%
  Yeah, the corporate interest aligns with testing perfectly, I think.
Because, as soon as you're doing any kind of maintenance,
it's great to have some confidence that you're not breaking everything,
and tests are essential for that.}{51}{4}{247}

\paragraphlf{\ref{subj:53} \--~Senior (10+) Retail}

\ievidence{%
Luckily for us there is the benefit that you can always learn
from the existing code and the existing
tests.
We try to be quite structured
And again, we name it like,
oh, you should do this,
given blah, blah - and hopefully these given cases
are like a bit exhaustive regarding your inputs.
We try to structure our tests the same way.
Given, when, then. First the variables,
then you set up your mocks, then you do the tests.
It all kind of follows.
Like, oh,
the rule is just you mock all external dependencies,
So you use, what's the correct term? stubs? mocks?
You don't have to mock it,
you can just fill the data objects.
That's the nice thing about it.
If you follow these best practices,
it almost writes itself.
}{53}{1}{726}

\ievidence{%
So we also do it a little bit [of testing]
like to gain this external confidence  .
I can't really say it's like 100\%,
like: ``oh, it's just the way our team likes it.''
There are also some of these requirements.
And again, especially with financial stuff
where it's these audits and everything,
there's a bit more pressure to at least write down
how we're doing this and this.}{53}{2}{927}

\ievidence{%
We really have to feel it is useful as a team.
Like this is these tests,
if we didn't feel they were useful, like the end-to-end tests,
they wouldn't be in the state they are in now,
but again, because it has helped us so many times
and we see it over and over again
when we develop those things,
we catch bugs like unit tests or whatever it didn't caught.
So it sounds like everyone has this sense
that it works and it's nice.}{53}{3}{938}

\paragraphlf{\ref{subj:55} \--~Senior (5-10) Software}

\ievidence{%
For me, it's more related to
how long you will maintain a project.
Because for example, for those client [non open source] projects,
if it's an easy project
that you will just develop
and you will deploy it to the client,
and you don't have much to do in terms of maintenance.
Then probably you might not need tests
at all because it's easy and you won't do changes on it.
And so you don't have much risk of
regressions.
Now, if it's a project with a lot of interactions
with different things are between
them and, and you have to, to keep whether:
It's [maintenance is] a project in its own,
so you have to keep
maintaining it and things like that.
Then you might need those tests to, to get to be sure.
}{55}{0.5}{209}

\ievidence{%
The benefit is clearly that if you have a test seat belt, basically, you,
you have less risk of having at least important bugs and important
regressions. For me, it's more about regressions.
Even more than bugs. And,
yes, you, it's more easy to be confident.
}{55}{0.75}{231}

\ievidence{%
when it's a bug where your people start to
lose some data and they try to save a document, for example, and they cannot, when they are
losing everything they typed, yeah, it's becoming really, really bad.
\faIcon{microphone}: Right. So what did you learn from it? Is there a test for this now?
\faIcon{comment}There is a test for it. And I actually, well, the bug here was quite complex, actually,
because it was the migration, it was database migration. So it was concerning multiple kinds
of database. Each time we do this kind of migration, we have to test on this different
database we support and stuff like that. And we have some strategies for that, but
it's not always working as we want. We cannot, well, we try to have automated tests
running on different database, but it's not always easy to write.
And yes, I wrongly tested that. And I was planning in January to provide another
migration inside the application. I decided to postpone the migration to take more time to test it.
Right? To \textit{ensure} it won't cause any more regression.
But yes, this kind of stuff
happens.
\faIcon{microphone}: Yeah, exactly. And you learned from it.
  \faIcon{comment}: Yeah.}{55}{1}{262}

\ievidence{%
Between the two teams,
the one working on the open source project
and the one working with
clients, you have actually the two views
because inside the client team,
you have this kind of
implicit knowledge and I'm always astonished. There's no documentation for that [testing strategies].
How do you know it?
\faIcon{microphone}: By sticking around? By putting your nose into different corners?
\faIcon{comment}: I don't like that. But yes.
I need to have a stuff explicit. Written and to be square.
\faIcon{microphone}: It helps, right?
\faIcon{comment}: Yeah.
\faIcon{microphone}: But then, maybe, if you do it for an internal team,
it also bothers some people.
\faIcon{comment}: Yeah, it takes time actually because you need to
ensure the rules evolved and to ensure everyone understands and agrees with them.}{55}{2}{356}

\paragraphlf{\ref{subj:57} \--~Senior (10+) Transport}

\ievidence{So that, that, that gives me, takes me three seconds.
And then when I figured out I will add it to the code.
So I'm iterating towards that.
But then for my say modules or classes to work, I usually write a unit test.
So I get that it's, it's very satisfying if you can get it to run under a second, just
also have that feedback.
And I don't try to think too much, like I'm not trying to read into the details too much.
I'm just like hammering and then I'm hitting the unit test and it's like, oh, it succeeded.
Well, okay.
Then the code is good.
Continue all that.
So you don't need to overburden yourself when you have such a fast system.}{57}{1}{355}

\ievidence{%
So I read that in a book and I thought it was so interesting, but I'm not really sure
how to apply it because I do want my regression tests even at some point in time.}{57}{1.5}{464}

\ievidence{I feel like people, teams or
  companies that don't have a good developer experience might
create that culture of like where nothing happens.
And that's dangerous as well.
So [testing] has some impact on that.
I don't know if this is true or not, but I guess companies can change teams or
like the code base might get swapped out, then you get new persons in -
and if there's no test on the code base - any sort of guardrails - if you're new to
it and you don't know the code base, then it's going to get really difficult to do anything
at all.
And those are situations that the company needs to consider,
but not really me as a
developer.
So that's why I also like this kind of things.
It's not testing anymore at this point,
or maybe it is? [thinking]
But that's why I really like this kind of very simple ideas which
have a big impact, like GitOps and automatic deployments.
Like, how do you deploy the code?
Well, I just push it.
If it passes a test, it gets deployed.
It also gives me the confidence if something fails that I can use Git, revert the commit
and push a new change and it will restore the situation to something that's
doable again.
It's also a saying: a bird on the branch is not afraid - it does not put its trust
in the branch, but rather on its wings.}{57}{2}{613}

\paragraphlf{\ref{subj:58} \--~Senior (10+) IT Services}

\ievidence{\faIcon{microphone}: So you said your customers start to hate you when you introduce bugs.
You must have said this because something happened. Right?
\faIcon{comment}: yeah, so personally, I broke [the software of] maybe the 20 biggest [customers].
And last year when I wasn't there - I was on vacation - we broke\ldots
let's say we had more customers when I started at the company.
And last year we completely broke [the software of] all of our customers. But really.
You know, basically it was a subtle bug caused by [one component] which generated some
non-valid code. We had a testing strategy,
but that was not perfect and there was a non-deterministic chance of something happening and it happened and
all customer applications crashed\ldots
It took something like an hour for the most impacted clients to recover\ldots
At least me and all the others that made the
armageddon-bug now understand that
it's completely normal to spend
99\% of the time on your pull request
to test your code and 1\% to add the code itself.
Because it's quite tricky to test.
Sometimes, your code involves web workers,
iframes a lot of things, you know?
Interaction with, code that may be running in the page of the customer.
So It's quite tricky, a good test,
but it's definitely worth doing
because For example when we introduced the big bug last year.
We spent More Than a month handling customers.
Sending the CTO, sending myself, discussing with customers.
So, it's better to invest in a propper CI/CD in the first place
with a proper testing strategy compared to wasting your time
basically convincing customers that it's not going to happen again.
\faIcon{microphone}: Because of the consequences?
\faIcon{comment}:
The consequences is customers
asking for money.
maybe they would say we want to X amount of money
but basically some customers say:
``Okay, because of a few customers were unable to buy anything
on a website for one hour. Usually we make this amount of money.
So you will pay us this we spend, this amount of money on ads during that time. so we don't want to pay you.''
These kinds of things.
Then the biggest one [consequence] is lack of trust,
lack of confidence and some renewals that may be more challenging.
So I don't think we lost anyone because of this
but I think we don’t have one more chance.
We did some mistakes in the past
that were smaller and we had another chance.
This one we felt like, this has to be the last time before
at least two or three years.
Otherwise, some big customers will turn away basically.}{58}{1}{00:17:21}

\ievidence{%
whenever someone joins on this components,
this [testing] is critical so we - Let's say I - will teach them to follow the process,
what could be the consequences.
We have post mortem for all this kind of things.
So, people can read internally.
We are really transparent.
\faIcon{microphone}: You answered the question already I guess but maybe let me make it very explicit: How do you trust yourself not to mess it up again?
\faIcon{comment}: The good thing is that we know why it happened and it was no one's fault.
Basically it was literally no one's fault.
It was something that had been there for maybe six years ago, even before I joined.
The non-reproducible build – we knew that we should have had reproducible builds.
We should have had better process, but it was never the priority until it became the priority.
It was just a lack of process. I mean, it was no one's fault and everyone's fault at the same time because
we all knew that some bug like this could happen. We didn't know it could come from non reproducible builds.
I mean, so no one felt guilty. No one was blamed because anyway, that's not how we operate.
So, it was easy.
Let's say everyone realized that now we operate at a huge scale so you need to test it properly.
}{58}{2}{00:21:12}

\ievidence{%
Sometime you need to have crisis to change your priorities
and basically in this case I wanted to invest more in JS topics,
but it's difficult to find a good Developer.
And we had other priorities but basically
we try to add as many Safeguard as possible.
Not only in the code. }{58}{3}{22:58}

\ievidence{You don't want to know whether today's code is working.
  You want to know that any kind of change that will
  be introducing a feature will
  be covered by the tests
  so you become a bit more paranoid.
  And yeah it really evolves over time.
We changed a lot.}{58}{4}{40:00:04}


\end{document}